\documentclass[aps,prd,preprint,groupedaddress]{revtex4}

\usepackage{epsfig}

\begin{document}

\def\gr{\tilde{G}}
\def\sq{\tilde{q}}
\def\gl{\tilde{g}}

\def\mr{m_{\tilde{G}}}
\def\ms{m_{\tilde{q}}}
\def\mi{m_{\tilde{q}_i}}
\def\mg{m_{\tilde{g}}}

\def\ts{t_{\tilde{q}}}
\def\ti{t_{\tilde{q}_i}}
\def\tg{t_{\tilde{g}}}
\def\us{u_{\tilde{q}}}
\def\ui{u_{\tilde{q}_i}}
\def\ug{u_{\tilde{g}}}

\def\d{{\rm d}}

\def\lp{\left. }
\def\rp{\right. }
\def\lr{\left( }
\def\rr{\right) }
\def\le{\left[ }
\def\re{\right] }
\def\lg{\left\{ }
\def\rg{\right\} }
\def\lb{\left| }
\def\rb{\right| }

\def\beq{\begin{equation}}
\def\eeq{\end{equation}}
\def\bea{\begin{eqnarray}}
\def\eea{\end{eqnarray}}

\preprint{hep-ph/0610160}
\preprint{LPSC 06-057}
\title{\vspace*{1mm}
 New Results for Light Gravitinos at Hadron Colliders -- \\
 Tevatron Limits and LHC Perspectives}
\author{Michael Klasen}
\email[]{klasen@lpsc.in2p3.fr}
\affiliation{Institut f\"ur Theoretische Physik, Universit\"at
 G\"ottingen, Friedrich-Hund-Platz 1, D-37077 G\"ottingen, Germany}
\affiliation{Laboratoire de Physique Subatomique et de Cosmologie,
 Universit\'e Joseph Fourier/CNRS-IN2P3,
 53 Avenue des Martyrs, F-38026 Grenoble, France}
\author{Guillaume Pignol}
\affiliation{Laboratoire de Physique Subatomique et de Cosmologie,
 Universit\'e Joseph Fourier/CNRS-IN2P3,
 53 Avenue des Martyrs, F-38026 Grenoble, France}
\begin{abstract}
We derive Feynman rules for the interactions of a single gravitino with
(s)quarks and gluons/gluinos from an effective supergravity Lagrangian
in non-derivative form and use them to calculate the hadroproduction cross
sections and decay widths of single gravitinos. We confirm the results
obtained previously with a derivative Lagrangian as well as those obtained
with the non-derivative Lagrangian in the high-energy limit and elaborate on
the connection between gauge independence and the presence of quartic
vertices. We perform extensive numerical studies of branching ratios, total
cross sections, and transverse-momentum spectra at the Tevatron and the LHC.
From the latest CDF monojet cross section limit, we derive a new and robust
exclusion contour in the gravitino-squark/gluino mass plane, implying that
gravitinos with masses below $2\cdot10^{-5}$ to $1\cdot10^{-5}$ eV are
excluded for squark/gluino-masses below 200 and 500 GeV, respectively.
These limits
are complementary to the one obtained by the CDF collaboration, $1.1\cdot
10^{-5}$ eV, under the assumption of infinitely heavy squarks and
gluinos. For the LHC, we conclude that SUSY scenarios with light gravitinos
will lead to a striking monojet signal very quickly after its startup.
\end{abstract}
\maketitle


\section{Introduction}
\label{sec:1}

Together with the possible existence of extra spatial dimensions,
supersymmetry (SUSY) remains the prime candidate for physics beyond the
Standard Model (SM). Among the many undisputed theoretical advantages of
the Minimal Supersymmetric SM (MSSM), the intimate connection of this new
space-time symmetry with electroweak symmetry breaking is of particular
importance. The search for SM or MSSM Higgs bosons as well as for spin-0
and spin-1/2 partners of the SM fermions and gauge bosons are therefore
often considered to be the most important tasks for present and future
collider experiments.

For many years, the focus has been on minimal supergravity (mSUGRA) models,
in which SUSY is broken by gravitational interactions and the lightest SUSY
particle (LSP) is the photino or, more generally, the lightest of four
neutralinos, $\tilde{\chi}^0_1$. Only around 1980 it was discovered
that the SUSY partner of the spin-2 graviton, the massless spin-3/2
gravitino, does not necessarily couple to matter with gravitational strength
only, but that its coupling can be enhanced to electroweak strength once
SUSY is broken through the super-Higgs mechanism and the associated
Goldstone fermion, the spin-1/2 goldstino, is absorbed to give the gravitino
its mass and its longitudinal degrees of freedom, making it the LSP
\cite{Fayet:1977vd,Casalbuoni:1988kv}.

The electroweak strength of goldstino interactions with massless photons and
photinos was then used to impose limits on the gravitino mass by comparing
total theoretical cross sections for electron-positron colliders to
experimental single-photon searches at PEP and PETRA, resulting in a first
mass limit of  $\mr\geq2.3\cdot10^{-6}$ eV \cite{Fayet:1986zc}.
Subsequently, the single-photon searches at LEP 1 and LEP 161 with cross
section limits of 0.1 and 1 pb implied gravitino masses above $10^{-3}$ and
$10^{-5}$ eV for light neutralinos of mass below 50 and 100 GeV,
respectively \cite{Lopez:1996ey}. These limits where, however, obtained
without imposing missing or observed photon energy cuts on the theoretical
cross section.

In 1988, the CDF collaboration published a cross section limit of 100 pb
for their monojet search at the Fermilab $p\bar{p}$ collider Tevatron
\cite{Abe:1988yj}, which they used to impose bounds on the squark-gluino
mass plane, but which could also be interpreted as the absence of a light
(s)goldstino signal, yielding $\mr>2.2\cdot10^{-5}$ eV and $\mg\geq100$ GeV
\cite{Dicus:1989gg}. This first hadron-collider analysis assumed, however,
very heavy squarks of $\ms\geq500$ GeV and was based on partonic
subprocesses involving only gluons and gluinos, but no (s)quarks. The
analysis was later re-applied to the 1996 CDF multijet cross section limit
of 1.4 pb \cite{Abe:1995pi}, yielding $\mr\geq3\cdot10^{-4}$ eV and
$\mg\geq200$ GeV \cite{Dicus:1996ua}. Predictions were also made for the
LHC, albeit for an assumed center-of-mass energy of 16 TeV
\cite{Drees:1990vj}.

While the gluon-gluon initial state dominates indeed for the production of
light final states at the LHC, it is well known that it is the
quark-antiquark luminosity that dominates at the Tevatron and that
quark-gluon initiated QCD Compton processes contribute significantly for
heavier final states at the LHC. A complete and robust study must therefore
take into account 1) all partonic subprocesses leading to the production of
single gravitinos, i.e.\ $q\bar{q}\to\gr\gl$, $gg\to\gr\gl$, and $qg\to\gr
\sq$, 2) the subsequent decay of the squark/gluino into an observed jet and
a second gravitino, i.e.\ $\gl\to\gr g$ and $\sq\to\gr q$, 3) up-to-date
collider energies, parton density functions (PDFs), values of $\Lambda_{\rm
QCD}$, and SUSY-breaking scenarios, 4) the experimental cuts on the jet and
missing transverse energies, and 5) the most recent experimental cross
section limits.

In Sec.\ \ref{sec:2}, we calculate analytical gravitino production cross
sections and decay widths using an effective supergravity Lagrangian in
four-component notation (see App.\ \ref{sec:a}) and the Feynman rules for
single gravitinos derived from it (see App.\ \ref{sec:b}). We discuss in
some detail the gauge-independence of our results and its relation to the
sign of interferences and the presence of quartic vertices. In Sec.\
\ref{sec:3}, we first present a concise review of gauge-mediated
SUSY-breaking (GMSB)
models, where gravitinos are naturally the lightest SUSY particles, and
discuss their implementation in different benchmark slopes. We then
establish the regions in which gluino/squark decays into gravitinos and jets
dominate. Next, we present the various subprocess contributions to the total
cross sections at the Tevatron and LHC and compute the jet and missing
transverse-momentum spectra, taking into account the gluino/squark decays.
Finally, we deduce a new limit on the gravitino mass from the latest CDF
monojet search and discuss the signal size and missing-$E_T$ trigger
thresholds at the LHC. Our conclusions are presented in Sec.\ \ref{sec:4}.
The discussion of cosmological constraints on the gravitino mass is beyond
the scope of this paper. For a recent analysis of Lyman-$\alpha$ forest and
WMAP data, assuming a light gravitino as a warm dark matter candidate in
GMSB models and yielding $\mr\leq16$ eV, we refer the reader to
\cite{Viel:2005qj} and the references therein.

\section{Analytical Results}
\label{sec:2}

In this Section, we present our analytical results for the hadroproduction
cross sections of single gravitinos with gluinos and squarks (Sec.\
\ref{sec:2a}) and the two-body decay widths of squarks and gluinos into
gravitinos with quarks and gluons (Sec.\ \ref{sec:2b}). They have been
obtained by using an effective supergravity Lagrangian in non-derivative
form (see App.\ \ref{sec:a}), from which the corresponding Feynman rules
(see App.\ \ref{sec:b}) have been derived.

\subsection{Production}
\label{sec:2a}

In $R$-parity conserving supersymmetry, single gravitinos can be produced
in strong interactions in association with either gluinos or squarks. In
addition, the associated production of gravitinos and gluinos proceeds
through two competing initial states, i.e.\ quark-antiquark or
gluon-gluon scattering, while gravitinos and squarks can only be produced
in quark-gluon scattering due to fermion number conservation. The
differential cross sections
\bea
 {\d\hat{\sigma}\over\d t} &=& {1\over2s}\,{1\over8\pi s}\,
 \overline{|M|}^2
\eea
depend in general on the SUSY particle masses, the usual Mandelstam
variables $s$, $t$, and $u$, and their mass-subtracted counterparts,
$t_{\sq,\gl}=t-m_{\sq,\gl}^2$ and $u_{\sq,\gl}=u-m_{\sq,\gl}^2$.
The gravitino mass $\mr$ will be neglected everywhere except in the
coupling constants, so that $t$ is integrated over the interval
$[-s+m_{\sq,\gl}^2\,;\,0]$.

\begin{figure}
 \centering
 \includegraphics[width=\columnwidth]{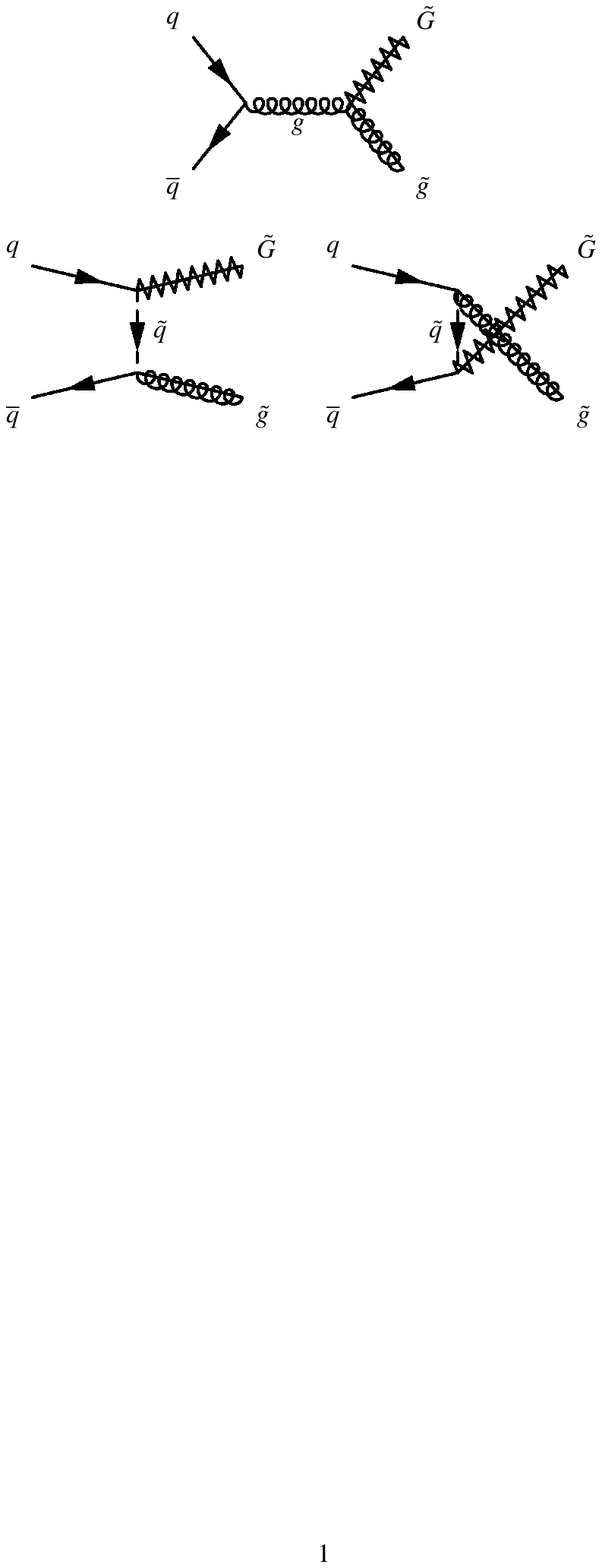}
 \caption{\label{fig:1}Leading-order Feynman diagrams for the
 production of a gravitino in association with a gluino in quark-antiquark
 collisions.}
\end{figure}

We first consider the process initiated by quarks and anti-quarks,
\beq
 q\bar{q}\to \gr\gl,
\eeq
whose contributing Feynman diagrams are shown in Fig.\ \ref{fig:1}.
%
%
The corresponding squared transition matrix element, averaged (summed)
over initial (final) state spins and colors and summed over left- and
right-handed squark exchanges,
\bea
 \overline{|M|}^2_{q\bar{q}\to\gr\gl} & = &
 {g_s^2 \, C_F \over 3 \,N_C \,M^2 \,\mr^2}\,
 \le {\mg^2\over s}\lr 2tu-\mg^2(t+u)\rr
 + \frac{m_{\tilde{q}}^4}{\ts^2} \,t\,\tg
 + \frac{m_{\tilde{q}}^4}{\us^2} \,u\,\ug \rp \nonumber\\
 & & \lp \hspace*{24mm}
 + \,{2\,\mg^2\,\ms^2\over\ts\,\us}\,\lr-2tu+\ms^2 (t+u)\rr \re,
 \label{eq:6}
\eea
is symmetric under the exchange of the $t$ and $u$ Mandelstam variables.
The $s$-channel contribution is individually gauge-independent, and the
$t$- and $u$-channel contributions are manifestly gauge-independent.
For gluino pair production, an interference term between the
$t$- and $u$-channel diagrams proportional to the squared gluino mass exists
\cite{Beenakker:1996ch}, but this term vanishes for gravitino-gluino
associated production linearly with the gravitino mass.

The contributions from individual diagrams that we obtain differ, of course,
from those presented in Eq.\ (4) of \cite{Kim:1997iw}, since our effective
Feynman rules are proportional to the SUSY particle masses, but our total
results agree. A related cross section has been computed with effective
Feynman rules in Eq.\ (28) of \cite{Lopez:1996ey} for the associated
production of gravitinos and neutralinos at lepton colliders. After
adjustment of masses, couplings, and color factors, it agrees with our
result. In the limit of negligible squark- and gluino masses, where the $t$-
and $u$-channel contributions vanish both due to the higher mass-dimension
of the squark coupling, our result agrees also with Tab.\ 1 in
\cite{Bolz:2000fu} when summed over left- and right-handed quarks. This
limit is, however, only applicable in the high-energy context of cosmology
\cite{Bolz:2000fu} and not at current hadron colliders.

\begin{figure}
 \centering
 \includegraphics[width=\columnwidth]{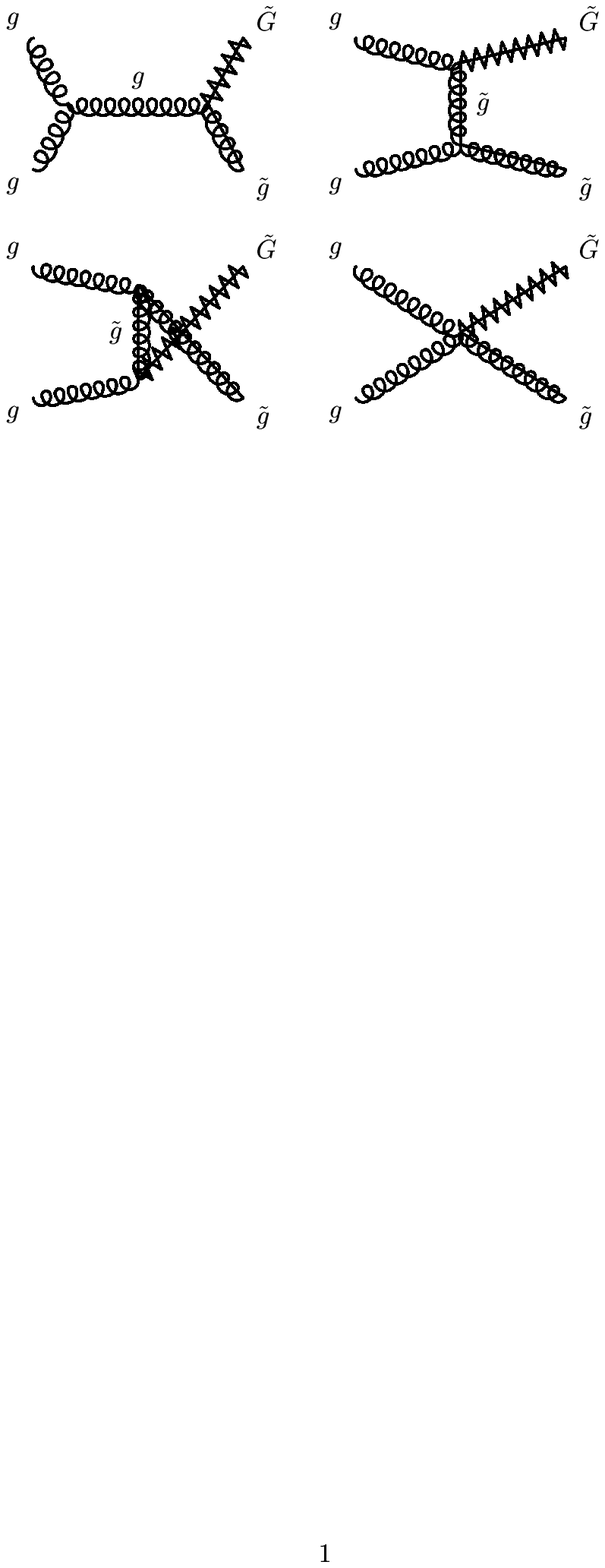}
 \caption{\label{fig:2}Leading-order Feynman diagrams for the
 production of a gravitino in association with a gluino in gluon-gluon
 collisions.}
\end{figure}

Next, we compute the competing gluon-initiated process
\beq
 gg\to \gr\gl,
\eeq
whose contributing Feynman diagrams are shown in Fig.\ \ref{fig:2}.
%
%
The gauge-independent total squared matrix element
\bea
 \overline{|M|}^2_{gg\to\gr\gl} & = &
 {g_s^2\,\mg^2 \over 6\,C_F\,M^2\,\mr^2}\,
 {s\,t\,u\over s^2\,\tg^2\,\ug^2}\,
 \le         tu\,(t^2+u^2)
    - \mg^2\,(t^3+6t^2u+6tu^2+u^3) \rp \nonumber \\ && \lp \hspace*{37mm}
    +2\mg^4\,(2t^2+7tu+2u^2)
    -5\mg^6\,(t+u)\re,
 \label{eq:8}
\eea
averaged (summed) over initial (final) state spins and colors, is again
symmetric under interchange of the final gravitino and gluino and
consequently also of the Mandelstam variables $t$ and $u$. Our result agrees
with Eq.\ (6) in \cite{Kim:1997iw} and also with Tab.\ 1 in
\cite{Bolz:2000fu} in the limit of small gluino mass. It also agrees with
Eq.\ (5) in \cite{Dicus:1996ua}, if its variables $t$ and $u$ are understood
to be their mass-subtracted counterparts and if its integration variable
$z$, whose definition is unfortunately missing, is assumed to be given by
$z=1+2t/(s-\mg^2)\,\in\,[-1;1]$.

\begin{figure}
 \centering
 \includegraphics[width=\columnwidth]{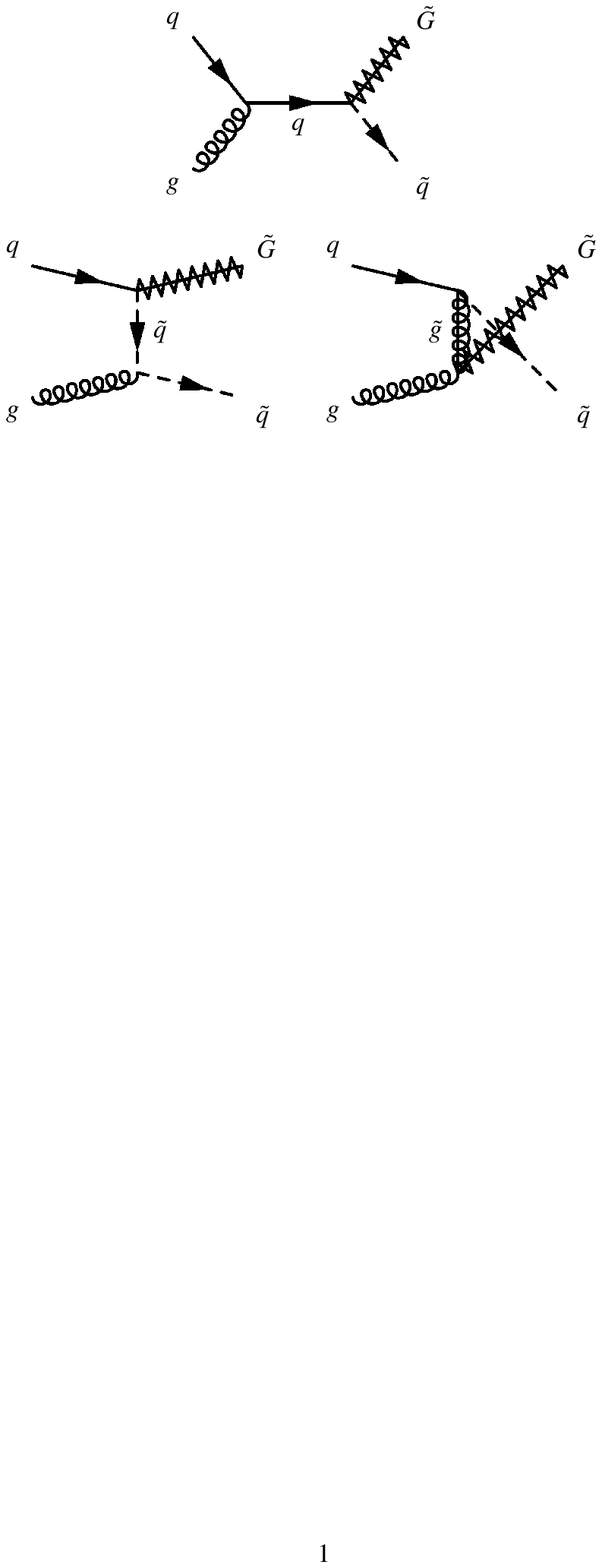}
 \caption{\label{fig:3}Leading-order Feynman diagrams for the
 production of a gravitino in association with a squark in quark-gluon
 collisions.}
\end{figure}

Finally, we analyze the associated production of gravitinos and squarks,
which is initiated by quark-gluon scattering,
\beq
 qg\to \gr\sq_i,
\eeq
and proceeds through the Feynman diagrams in Fig.\ \ref{fig:3}.
%
%
In this case, the squared matrix element for a squark of a given chirality
$i$ is
\bea
 \overline{|M|}^2_{qg\to\gr\sq_i} & \!\!=\!\! &
 {g_s^2 \over 12 N_C M^2 \mr^2}
 \le {\mi^4\over s\ti^2}\,(-u)\,(t^2+\mi^4)
 \!+\!{\mg^2\over\ug^2}\,(-u)\,(tu+\mg^2s)
 \!+\!{\mg^2\,\mi^4\over\ti\,\ug}\,(2u)\re\!,~~
 \label{eq:10}
\eea
where the first term in the squared brackets denotes the gauge-independent
sum of $s$- and $t$-channel contributions including their interference,
while the $u$-channel contribution in the second term is individually
gauge-independent. The third term corresponds to the gauge-independent 
sum of $s$- and $t$-channel interferences with the $u$-channel. The same
result is obtained for the charge-conjugated process,
\beq
 \bar{q}g\to \gr\sq_i^*.
\eeq
This leads to a factor of two for $p\bar{p}$ colliders with a neutral
initial state such as the Tevatron, but not for $pp$ colliders such as the
LHC, where the parton densities are not charge-symmetric. Our result agrees
with the one in \cite{Kim:1997iw}, which has been obtained
with derivative couplings including a quark-gluon-gravitino-squark vertex
contribution. Note that this quartic vertex is absent in the effective
theory \cite{Lee:1998aw}, since it would spoil the gauge-independence. At
high energies, only the $u$-channel contribution survives, so that our
result agrees with the one in Tab.\ 1 of \cite{Bolz:2000fu}.

\subsection{Decay}
\label{sec:2b}

\begin{figure}
 \centering
 \includegraphics[width=0.49\columnwidth]{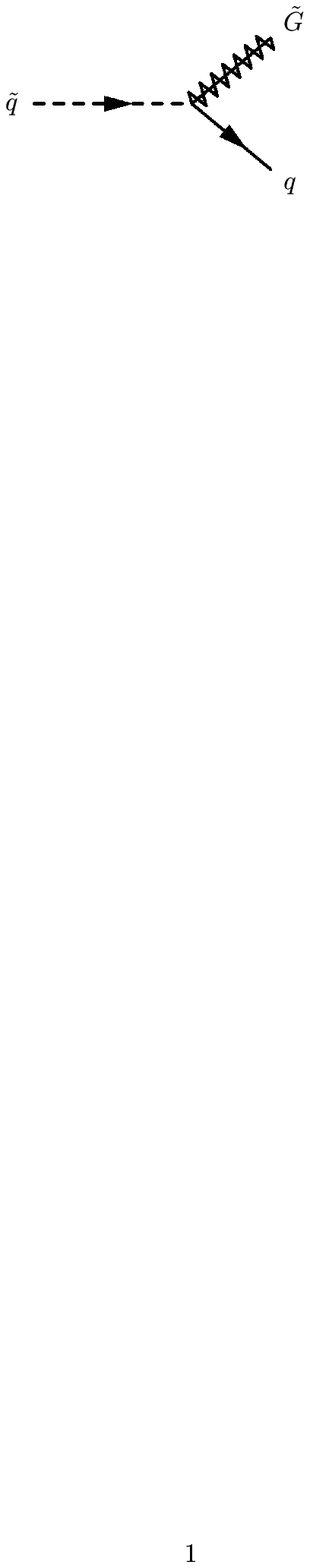}
 \includegraphics[width=0.49\columnwidth]{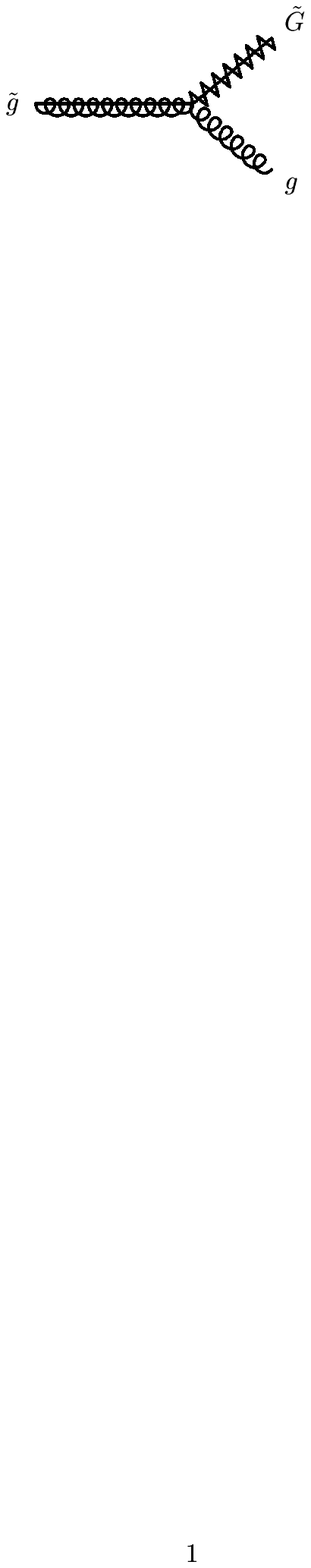}
 \caption{\label{fig:4}Leading-order Feynman diagrams for the decay of a
 squark into a gravitino and a quark (left) and a gluino into a gravitino
 and a gluon (right).}
\end{figure}

Heavy squarks and gluinos may decay either directly or through cascades
into the lightest SUSY particle, which we assume to be the gravitino.
Direct decays, which dominate for light gravitinos \cite{Dicus:1989gg},
proceed through the Feynman diagrams shown in Fig.\ \ref{fig:4}, and the
corresponding partial widths
%
%
\bea
 {\d\Gamma\over\d t} &=& {1\over2m_{\sq,\gl}}\,{1\over8\pi m_{\sq,\gl}^2}\,
 \overline{|M|}^2
\eea
are obtained from the squared transition matrix elements after integration
of the Mandelstam variable $t$ over the interval $[-m_{\sq,\gl}^2+m_{q,g}^2
\,;\,0]$. For a squark of a given chirality $i$, the squared matrix element
\bea
 \overline{|M|}^2_{\sq_i\to\gr q} & = &
 {(\mi^2-m_q^2)^3\over 3 \, M^2 \, \mr^2}
\eea
leads then to the partial width
\bea
 \Gamma_{\sq_i\to\gr q} & = &
 {\mi^5 \over 48 \, \pi \, M^2 \, \mr^2}\,
 \lr 1-{m_q^2\over\mi^2}\rr^4.
\eea
Since the gluon mass is, of course, zero $(m_g=0)$, the squared gluino decay
matrix element, averaged (summed) over initial (final) spins is
\bea
 \overline{|M|}^2_{\gl\to\gr g} & = &
 {\mg^6 \over 3 \, M^2 \, \mr^2},
\eea
leading to the partial width
\bea
 \Gamma_{\gl\to\gr g} & = &
 {\mg^5 \over 48 \, \pi \, M^2 \, \mr^2}.
\eea
These results are well-known \cite{Ambrosanio:1996jn}. They agree, in
particular, with the general result in Eq.\ (6.24) of \cite{Martin:1997ns},
valid for the decay of any heavier SUSY particle into its Standard Model
partner and a lighter gravitino.

\section{Numerical Results}
\label{sec:3}

\subsection{Gauge Mediated Supersymmetry Breaking}
\label{sec:3a}

In Gauge Mediated Supersymmetry Breaking (GMSB) models, SUSY
breaking occurs in a secluded sector at the scale $\langle F \rangle$,
related to the gravitino mass by $\mr=\langle F \rangle/(\sqrt{3}M)$, and
is transmitted to the observable sector by a chiral superfield $S$ and
$n_q$ quark-like and $n_l$ lepton-like messenger fields
\cite{Giudice:1998bp,Martin:1997ns}. The superfield $S$
is a gauge-singlet, but its scalar and auxiliary components overlap with the
gravitino and acquire vacuum expectation values $\langle S \rangle$ and
$\langle F_S \rangle$. The messenger fields then acquire a mass $M_{\rm
mess} \simeq\langle S\rangle$ through Yukawa couplings to the superfield
$S$. They are given the same Standard Model gauge couplings to the
observable fields as ordinary quarks and leptons, so that they can induce
gaugino and sfermion masses through one- and two-loop self-energy diagrams,
respectively. The lightest SUSY particle is always the gravitino, and it is
for this reason, that we concentrate our numerical study of gravitino
production at hadron colliders on GMSB scenarios.

Besides $M_{\rm mess}$, $n_q$, and $n_l$, GMSB scenarios are determined by
the ratio of Higgs vacuum expectation values, $\tan\beta$, the 
sign of the Higgs mass-parameter $\mu$, and by the auxiliary vacuum
expectation value, $\langle F_S \rangle$, which is related to the mass
splitting of the messenger fields and realistically considerably smaller
than both the squared mass scale of the messenger fields, $\langle S
\rangle^2$, and the fundamental SUSY-breaking scale, $\langle F\rangle$.
It is usually re-expressed in terms of an effective SUSY-breaking scale,
$\Lambda=\langle F_S\rangle/\langle S\rangle$.

\begin{figure}
 \centering
 \includegraphics[width=0.9\columnwidth]{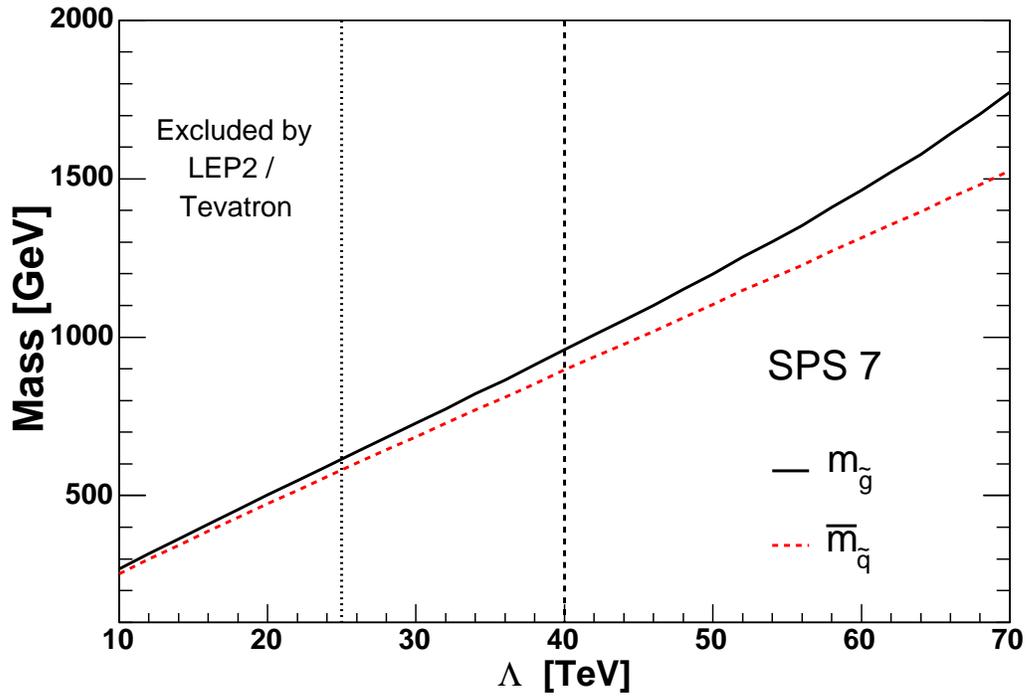}
 \includegraphics[width=0.9\columnwidth]{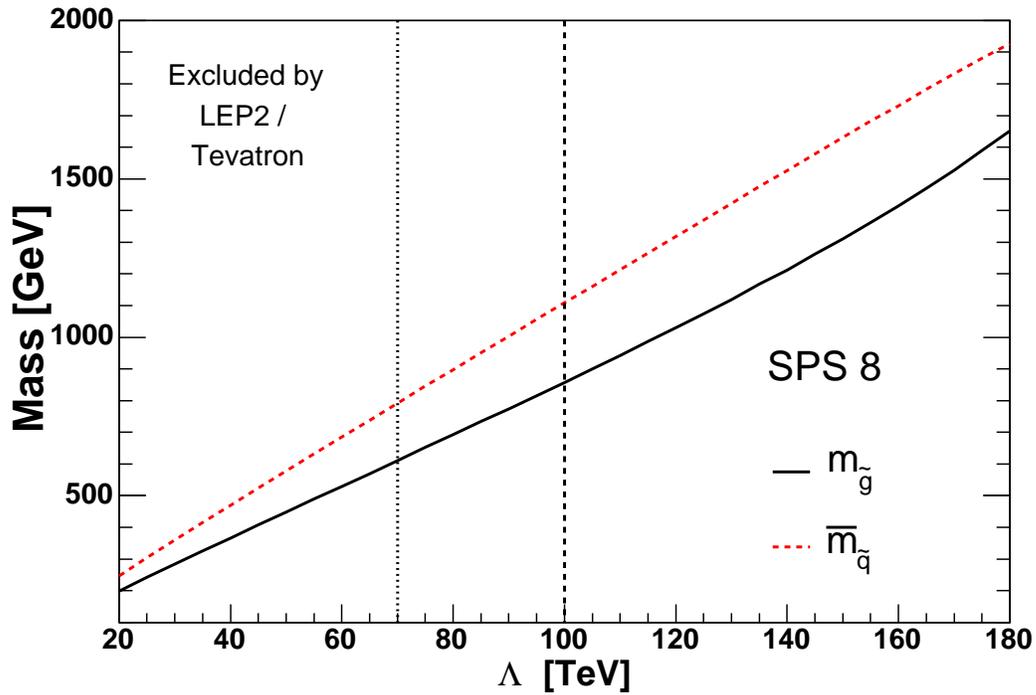}
 \caption{\label{fig:5}Average squark and physical gluino masses for the
 GMSB benchmark slopes SPS 7 (top) and SPS 8 (bottom) as a function of the
 effective SUSY-breaking scale $\Lambda$, with $M_{\rm mess}/\Lambda=2$,
 $\tan\beta=15$, and $\mu>0$ fixed.}
\end{figure}

A number of SUSY benchmark scenarios have been proposed in
\cite{Allanach:2002nj} in order to facilitate detailed comparisons between
SUSY searches at different colliders and with different signals/backgrounds.
In particular, we show in Fig.\ \ref{fig:5} the average squark and physical
%
%
gluino masses for the two GMSB scenarios proposed in \cite{Allanach:2002nj},
SPS 7 and 8, as a function of the effective SUSY-breaking scale $\Lambda$,
with $M_{\rm mess}/\Lambda=2$, $\tan\beta=15$, and $\mu>0$ fixed. For SPS 7,
the benchmark point (indicated by a vertical dashed line) is at $\Lambda=40$
TeV and $n_q=n_l=3$, leading to a stau $(\tilde{\tau}_1)$ next-to-lightest
SUSY particle (NLSP), while for SPS 8, the benchmark point is at
$\Lambda=100$ TeV and $n_q=n_l=1$, leading to a neutralino
($\tilde{\chi}^0_1$) NLSP. The regions that have already been excluded by
LEP2 and Tevatron searches for light neutralinos and charginos in GMSB
scenarios lie to the left of the vertical dotted line \cite{Yao:2006px}.
The physical masses in Fig.\ \ref{fig:5} have been obtained by imposing
boundary conditions at the Grand Unification Theory (GUT) scale and evolving
them to the electroweak symmetry breaking scale via renormalization group
equations using the computer program {\tt SUSPECT} \cite{Djouadi:2002ze}.
Note that the mass hierarchy of gluinos and squarks at SPS 7, $\mg\geq
\overline{m}_{\sq}$, is reversed at SPS 8, where $\mg\leq
\overline{m}_{\sq}$.

\subsection{Branching Ratios}
\label{sec:3b}

We are now in a position to determine the regions in SUSY parameter space,
where the squarks and gluinos, that are produced in association with the
gravitino at hadron colliders, decay dominantly into a two-body final state
with a second gravitino and a quark or gluon, leading to an experimentally
identifiable monojet signal with large missing transverse energy.

\begin{figure}
 \centering
 \includegraphics[width=0.9\columnwidth]{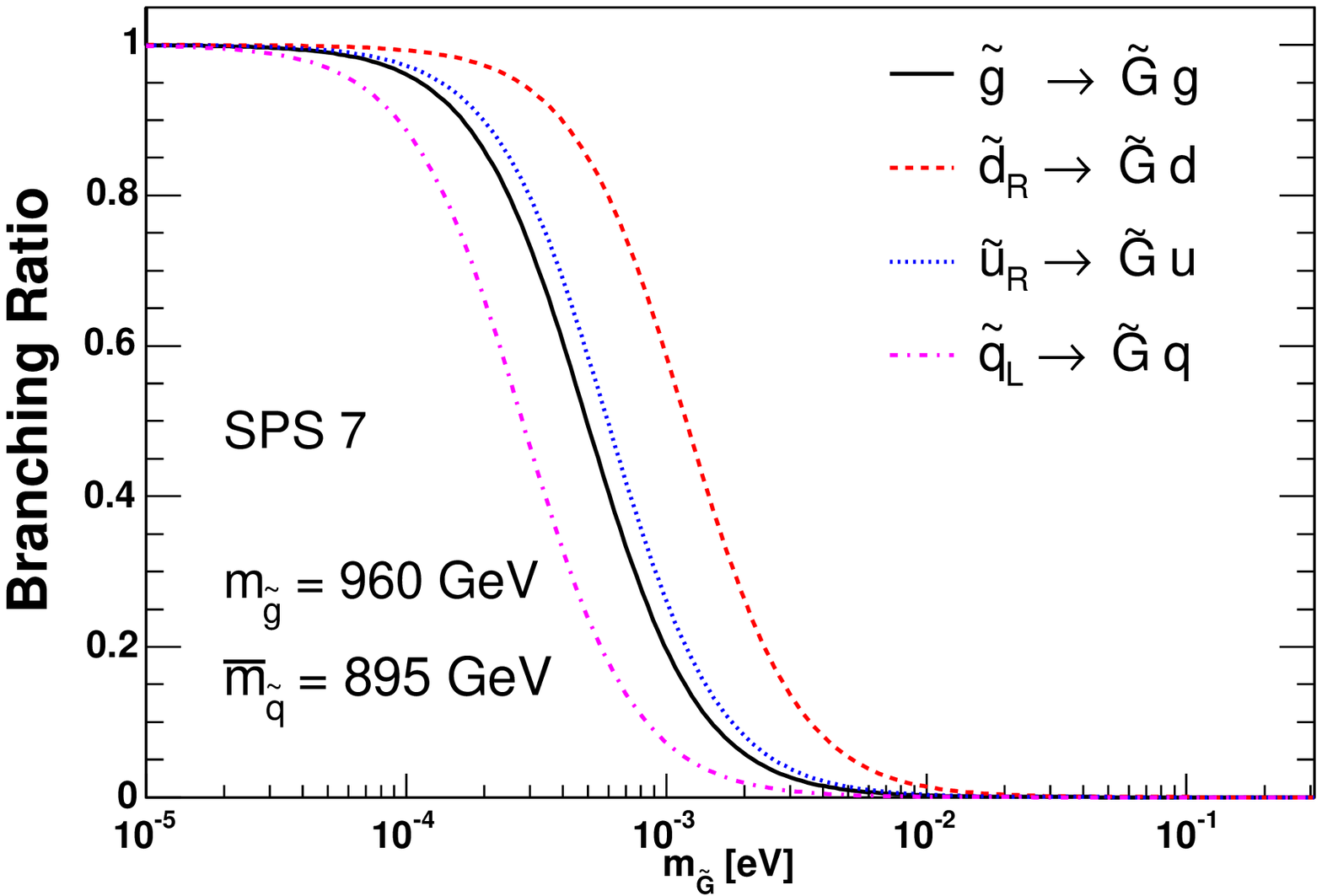}
 \includegraphics[width=0.9\columnwidth]{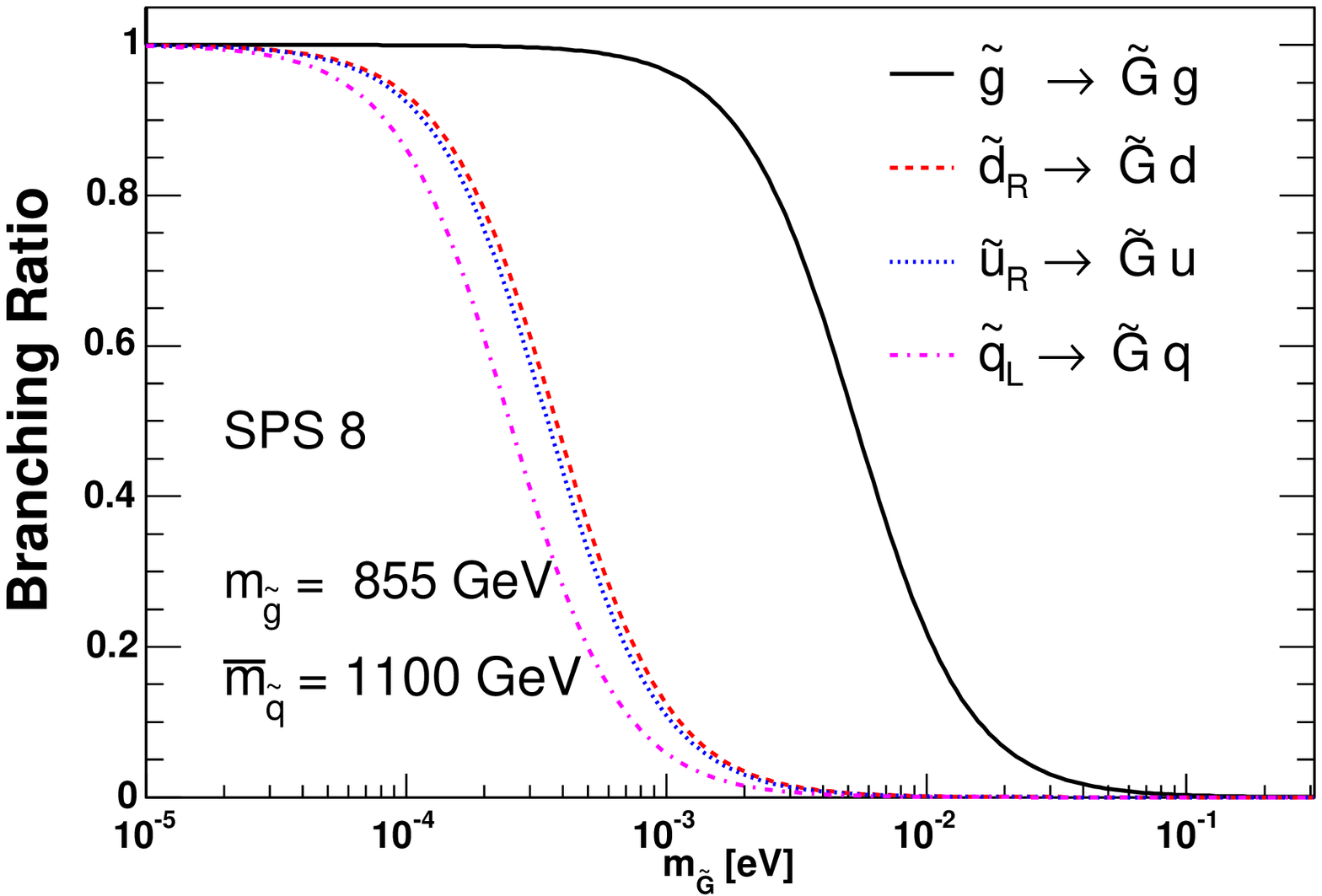}
 \caption{\label{fig:6}Branching ratios of gluinos/squarks into gravitinos
 and gluons/quarks as a function of the gravitino mass with the other SUSY
 masses fixed to their SPS 7 (top) and SPS 8 (bottom) GMSB benchmark
 values.}
\end{figure}

\begin{figure}
 \centering
 \includegraphics[width=0.9\columnwidth]{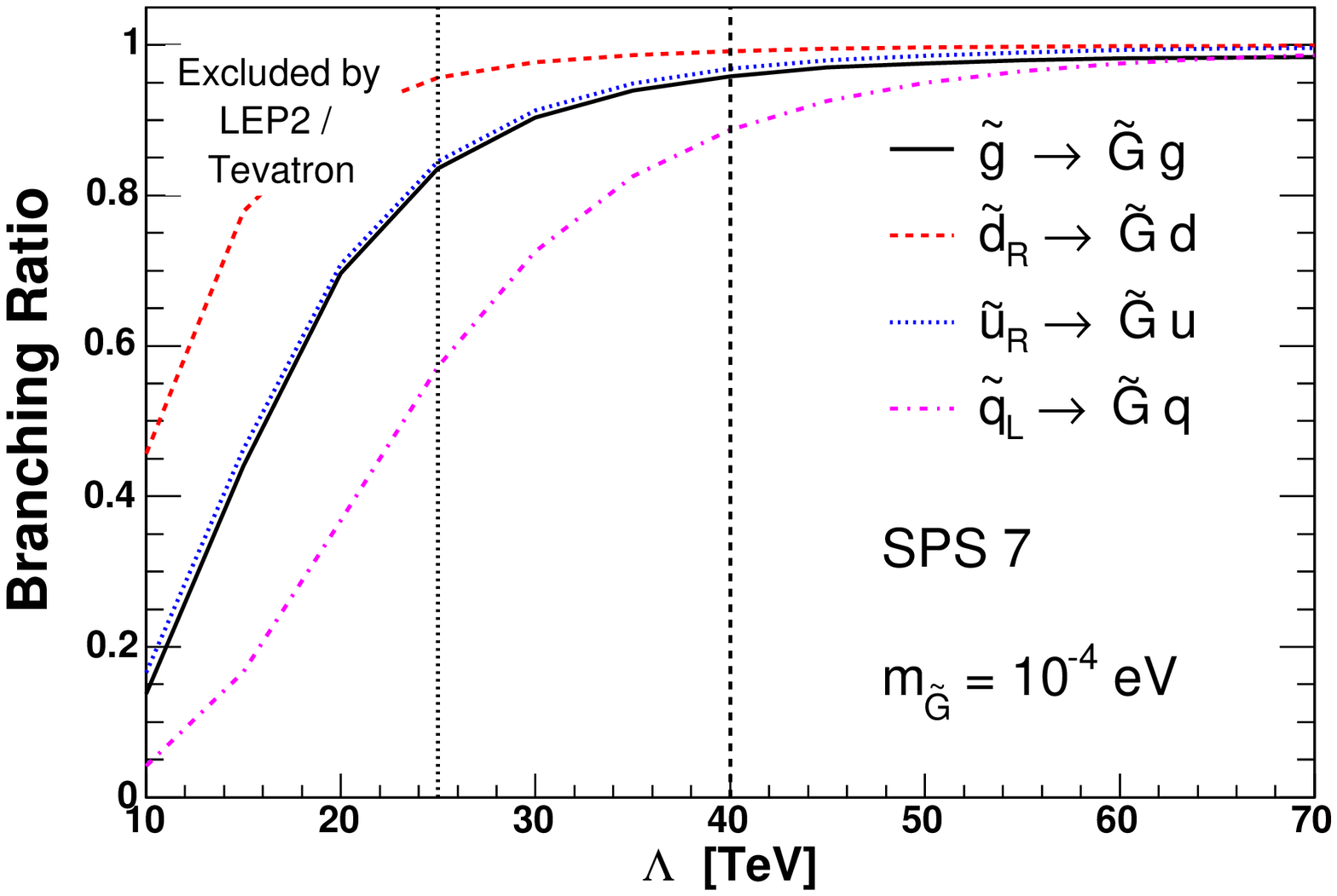}
 \includegraphics[width=0.9\columnwidth]{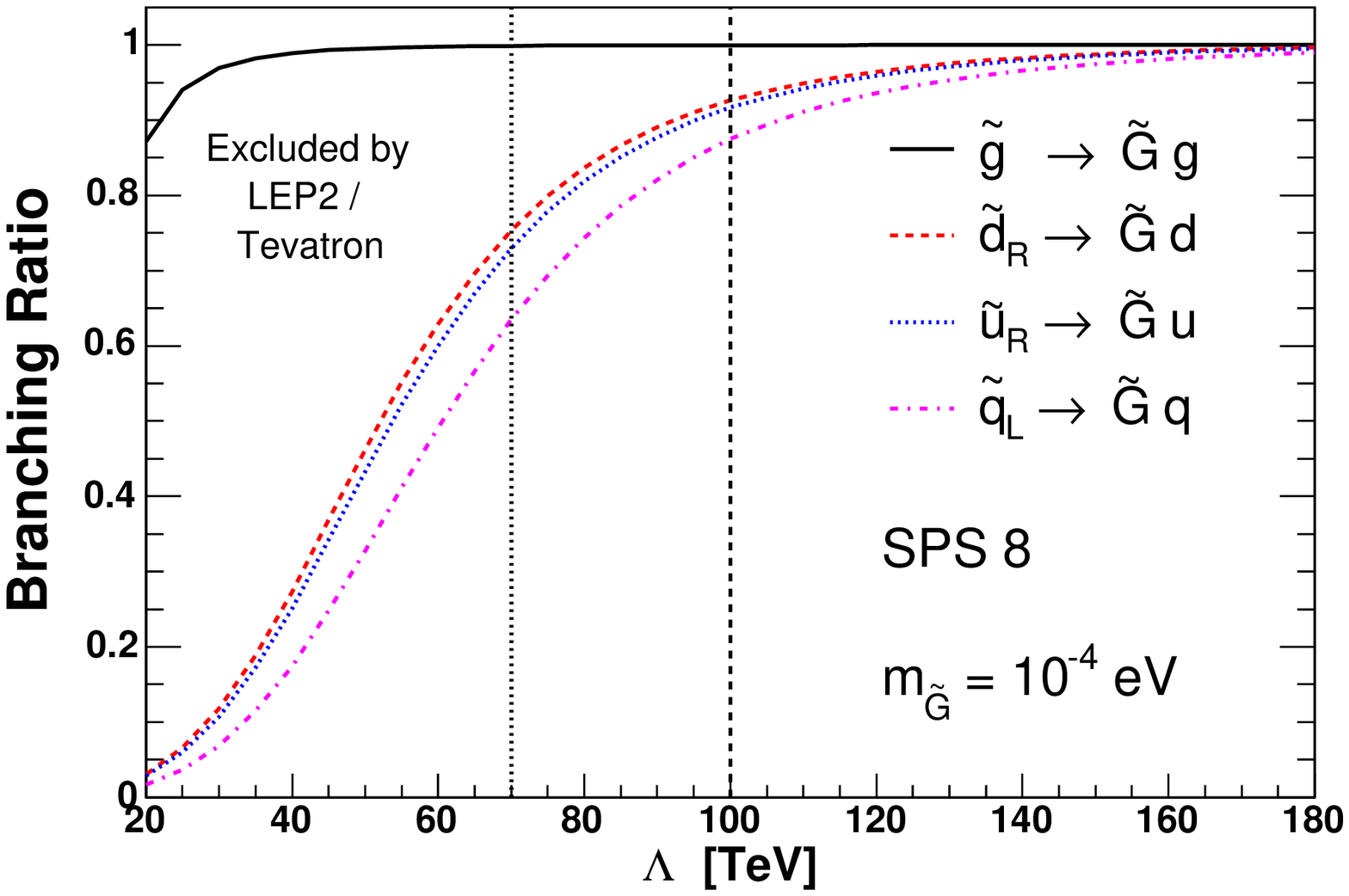}
 \caption{\label{fig:7}Branching ratios of gluinos/squarks into gravitinos
 and gluons/quarks for $\mr=10^{-4}$ eV and the GMSB benchmark slopes SPS 7
 (top) and SPS 8 (bottom) as a function of the effective SUSY-breaking scale
 $\Lambda$.}
\end{figure}

To this end, we evaluate the decay widths $\Gamma_{\gr}$ calculated in Sec.\
\ref{sec:2b} in the GMSB scenarios discussed in Sec.\ \ref{sec:3a} and
compare them to the competing total decay width $\Gamma_{\rm MSSM}$ of
gluinos and squarks into MSSM two-body final states up to one-loop level
and those into three- and four-body final states at tree-level as
implemented in the computer program {\tt SDECAY} \cite{Muhlleitner:2003vg}.
The resulting branching ratios
\bea
 {\rm BR}&=&{\Gamma_{\gr}\over\Gamma_{\rm MSSM}+\Gamma_{\gr}}
\eea
of gluinos and squarks are shown in Figs.\ \ref{fig:6} and \ref{fig:7} as
%
%
a function of the gravitino mass and of the effective SUSY-breaking scale
%
%
$\Lambda$, defining the GMSB benchmark slopes SPS 7 (top) and SPS 8
(bottom).

For both benchmark points, we observe that left- and right-handed, up- and
down-type squarks decay dominantly with BR $\geq 0.9$ into gravitinos, if
the gravitino mass does not exceed $\mr\leq10^{-4}$ eV. While the
SUSY-breaking scale $\Lambda$ must not lie significantly below the benchmark
points of $\Lambda=$ 40 and 100 TeV, respectively, these regions are already
largely excluded by LEP2 and Tevatron searches for light neutralinos and
charginos in GMSB scenarios \cite{Yao:2006px}.

For gluinos, the conclusions are quite similar for SPS 7, but more
optimistic for SPS 8 with the decay into gravitinos dominating up to
$\mr\leq10^{-3}$ eV for all physical values of $\Lambda$. Above these
limits, the decay chains depend essentially on the mass hierarchy of the
SUSY spectrum with $\gl\to\sq q$ and $\sq\to\tilde{\chi}_i^{0,\pm}
q^{(\prime)}$ at SPS 7, whereas $\sq\to\gl q$ and $\gl\to
\tilde{\chi}_i^{0,\pm}q \bar{q}^{(\prime)}$ at SPS 8, leading
in general to more complicated multijet signals. Note that all of these
decays are instantaneous with decay lengths around or below 1 fm, so that
they occur close to the primary vertex and well inside any collider
detector.

While our numerical results for squark decays are new, gluino decays have
been studied quite some time ago in a simple SUSY scenario with a massless
photino NLSP. In Fig.\ 1 of \cite{Dicus:1989gg}, the two-body decay $\gl\to
\gr g$ has been compared with the tree-level three-body decay $\gl\to
\tilde{\gamma}q\bar{q}$, neglecting all other decay modes and for $\mg=100$
GeV and $\ms=500$ GeV. The conclusion there was that BR $\geq 0.9$ up to
$\mr\leq10^{-4}$ eV, which compares quite favorably with our result at SPS
8 and $\Lambda=20$ TeV (see the lower part of Fig.\ \ref{fig:7}), where the
masses $\mg=200$ GeV, $\overline{m}_{\sq}= 250$ GeV, and $m_{\tilde{\chi}
^0_1}=14$ GeV are of similar magnitude. For $\ms=1000$ GeV, the gravitino
decay mode was found to dominate up to $\mr=5\cdot 10^{-3}$ eV (see Fig.\ 1
of \cite{Dicus:1996ua}). This compares again favorably with our result at
the benchmark point SPS 8 (see lower part of Fig.\ \ref{fig:6}), where
$\overline{m}_{\sq}= 1100$ GeV. Related results for a massless photino NLSP
and gluino masses between 200 and 750 GeV and squark masses between 500 and
2000 GeV can furthermore be found in Fig.\ 1 of \cite{Drees:1990vj}.

\subsection{Tevatron}
\label{sec:3c}

The total hadronic cross section for gravitino-gluino or gravitino-squark
associated production
\bea
 \sigma &~=&
 \int_{m^2/S}^1\!\d\tau\!\!
 \int_{-1/2\ln\tau}^{1/2\ln\tau}\!\!\d y
 \int_{t_{\min}}^{t_{\max}} \d t \ \sum_{a,b}\
 f_{a/A}(x_a,M_a^2) \ f_{b/B}(x_b,M_b^2) \ {\d\hat{\sigma}_{ab}\over\d t}
\eea
can be obtained by convolving the partonic cross sections
d$\hat{\sigma}_{ab}/$d$t$ presented in Sec.\ \ref{sec:2a} with the parton
density functions (PDFs) $f_{a,b/A,B}$ at the factorization scale $M_{a,b}$.
Since the PDFs vanish rapidly, as the longitudinal momentum fractions
$x_{a,b}$ of the partons $a,b$ in the external hadrons $A,B$ approach unity,
the available partonic center-of-mass energy $s=x_ax_bS$ represents only a
fraction $\tau=x_ax_b$ of hadronic center-of-mass energy $S$, and the
experimentally accessible mass range for searches of new SUSY particles is
naturally limited. We consider the initial gluons and five light quarks to
be massless and denote the average final state mass by $m$. At the LHC,
both $A$ and $B$ represent protons, which will collide with $\sqrt{S}=14$
TeV starting in 2008, whereas at the Tevatron, $B$ represents an anti-proton
beam with $\sqrt{S}=1.8$ TeV at the completed Run I and 1.96 TeV at the
current Run II.

Numerical predictions for single-gravitino hadroproduction cross sections at
Run I of the Tevatron have been presented in Figs.\ 4, 5, and 6 of
\cite{Kim:1997iw} as a function of $\mg$ and $\ms$, respectively. We have
verified these results by fixing the strong coupling to its world average
value $\alpha_s(M_Z)=0.118$ and convolving our partonic cross sections
in Sec.\ \ref{sec:2a} with the (nowadays obsolete) set of PDFs of
\cite{Rangarajan:Private}, evolved from the starting scale $Q_0=2$ GeV and
using a value of $\Lambda_{\rm LO}^{n_f=5}=144$ MeV to the factorization
scale $M_a=M_b=m_{\gl,\sq}$. In the following, we will, however, use the
modern PDFs of CTEQ6L1 \cite{Pumplin:2002vw}, which correspond to a one-loop
running of the strong coupling $\alpha_s(\mu)=g_s^2/(4\pi)$ and a QCD scale
parameter of $\Lambda^{n_f=5}_{\rm LO}=165$ MeV, derived from the
world-average value of $\alpha_s(M_Z)=0.118$ \cite{Yao:2006px}. The
renormalization scale $\mu$ and the factorization scales $M_{a,b}$ will be
fixed to the average particle mass $m=(\mr+m_{\tilde{q},\tilde{g}})/2$ in
the final state.

\begin{figure}
 \centering
 \includegraphics[width=0.9\columnwidth]{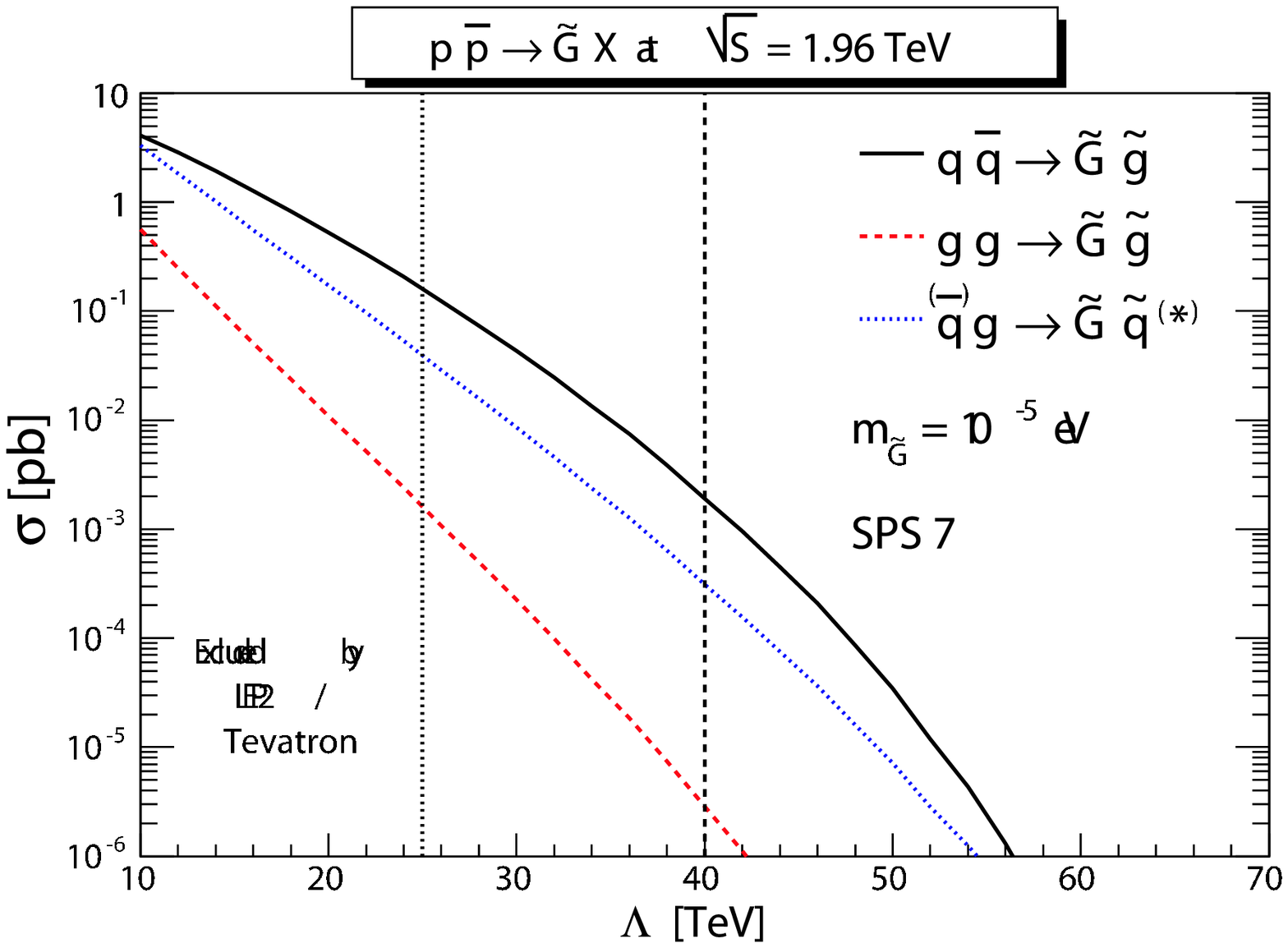}
 \includegraphics[width=0.9\columnwidth]{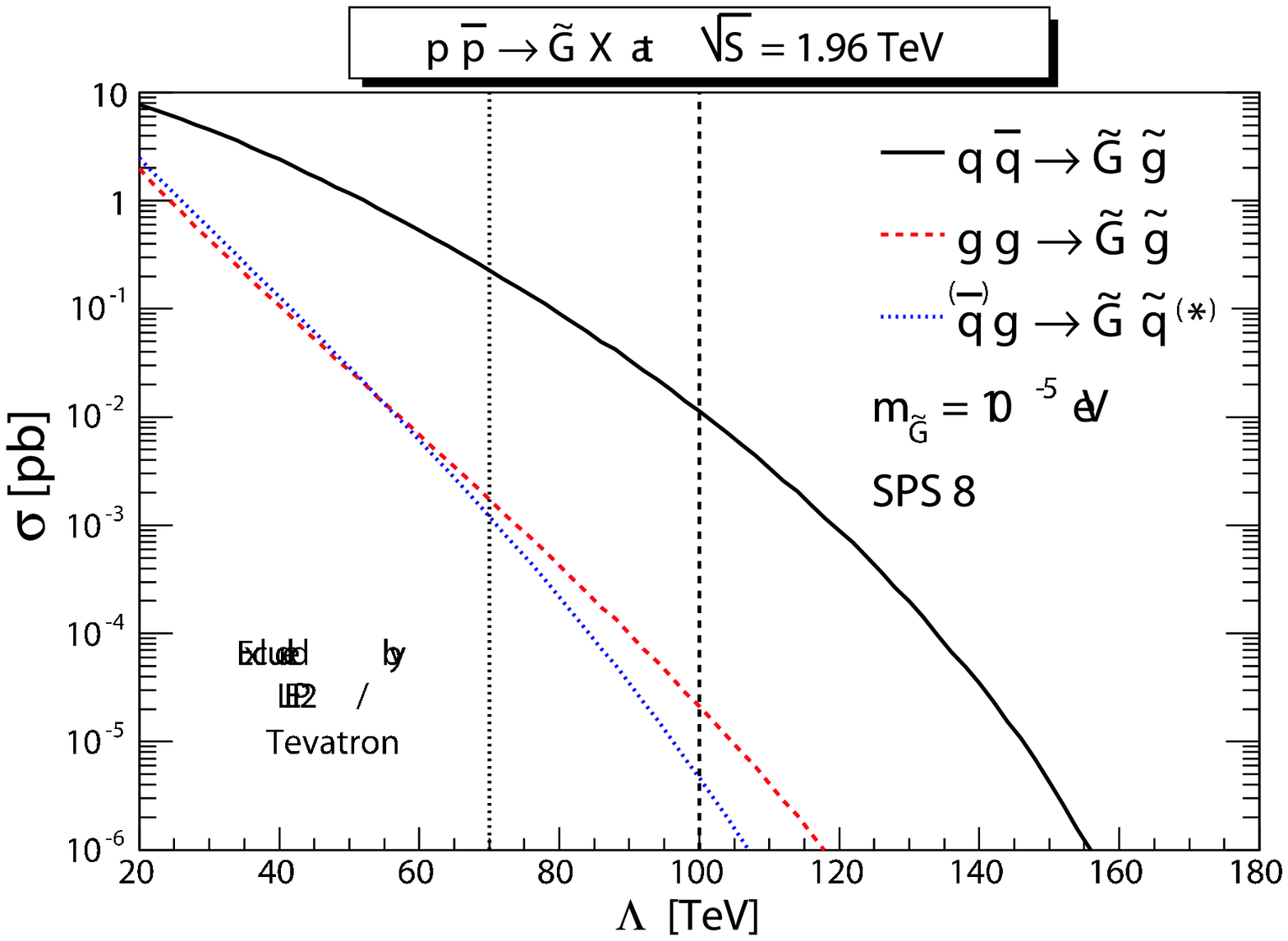}
 \caption{\label{fig:8}Total cross sections of gravitino and gluino/squark
 associated production at Run II of the Tevatron for $\mr=10^{-5}$ eV and
 the GMSB benchmark slopes SPS 7 (top) and SPS 8 (bottom) as a function of
 the effective SUSY-breaking scale $\Lambda$.}
\end{figure}

In Fig.\ \ref{fig:8}, the total cross section of the associated production
%
%
of gravitinos and gluinos or squarks is shown for the GMSB benchmark
scenarios SPS 7 (top) and 8 (bottom) and a gravitino mass of $\mr=10^{-5}$
eV as a function of the effective SUSY-breaking scale $\Lambda$. While
the quark-gluon initiated production of gravitinos and squarks can
contribute significantly to the total event sample for SPS 7, where
$\overline{m}_{\sq}\leq\mg$, the largest contribution at SPS 7, and even
more so at SPS 8, where $\mg\leq\overline{m}_{\sq}$, comes from the
subprocess $q\bar{q}\to\gr\gl$. This is of course due to the large
quark-antiquark luminosity at the Tevatron. At Run II, where the integrated
luminosity has already reached 2.6 fb$^{-1}$ and is expected to increase to
4.4 to 8.8 fb$^{-1}$ until the final shutdown in 2009
\cite{Shiltsev:2004uc}, the CDF and D0 experiments should be able to
discover light gravitinos with masses up to 10$^{-5}$ eV in events with a
single jet and large missing transverse energy for values of $\Lambda$ above
the current exclusion limits (25 and 70 TeV for SPS 7 and 8, respectively).
If we assume the large-$E_T$ monojet signal to have very little SM
background (see Fig.\ \ref{fig:9} below) and thus to be experimentally
identifiable with high efficiency,
we can define the visible region by the point where the total cross section
falls to 1 fb, so that only a few events will be recorded. The discovery
reach then extends up to the benchmark point (40 TeV) at SPS 7, where
$\mg=950$ GeV and $\overline{m}_{\sq}=890$ GeV, and even up to 120 TeV at
SPS 8, where $\mg=1000$ GeV and $\overline{m}_{\sq}=
1300$ GeV. For lighter gravitino masses, the total cross section scales
trivially according to Eqs.\ (\ref{eq:6}), (\ref{eq:8}), and (\ref{eq:10}),
i.e.\ with the inverse of the squared gravitino mass.

For a detailed, model-independent account of the experimentally identifiable
monojet signal of the associated production of a gravitino with a squark or
gluino, we have to include the decay of the latter into a second gravitino
and an observed jet. If we continue to neglect $\mr$, except in the
coupling, the cross section for a massless three-body massless final state
can be written as
\bea
 \d\sigma & \!=\! &
 {1 \over 2 s} \int
 f_{a/A}(x_a,M_a^2) \,\d x_a\,
 f_{b/B}(x_b,M_b^2) \,\d x_b\,
 \overline{|M|}^2_{2\to3} \,
 (2\pi)^{4} \, \delta^{4}\left(p_a+p_b-\sum_{i=1}^3 p_i\right)\,
 \prod_{i=1}^{3} {\d^3 p_i\over (2\pi)^3 \,2 E_i}\nonumber\\
 & \!=\! &
 {1 \over 2 s} \int
 f_{a/A}(x_a,M_a^2) \,
 f_{b/B}(x_b,M_b^2) \,
 \overline{|M|}^2_{2\to3} \,
 (2\pi)^{-5} \,
 {p_{T_1}\over 2}\d p_{T_1} \d\eta_1\d\phi_1
 {p_{T_2}\over 2}\d p_{T_2} \d\eta_2\d\phi_2
 {1 \over S}\d\eta_3,
\eea
where $p_{T_1}$ represents the observed jet transverse momentum, which is
balanced by the missing transverse momentum of the two gravitinos. Since
the squark or gluino decay width is of the order of $\hbar c~/~1$ fm = 0.2
GeV (see Sec.\ \ref{sec:3b}), we can apply the narrow-width approximation
to rewrite the squared and averaged $2\to3$ scattering matrix element as 
\bea
 \overline{|M|}^2_{2\to3} &=&
 \overline{|M|}^2_{2\to2} \,
 \left|{1\over s_{12}-m_{\sq,\gl}^2+im_{\sq,\gl}\Gamma_{\sq,\gl}}\right|^2\,
 |M|^2_{1\to2} \nonumber \\
 &\to&
 \overline{|M|}^2_{2\to2} \,
 {\pi\,\delta(s_{12}-m_{\sq,\gl}^2)\over m_{\sq,\gl}\,\Gamma_{\sq,\gl}}\,
 |M|^2_{1\to2}.
\eea
By fixing the azimuthal angle of the observed jet to $\phi_1=0$, the squared
invariant mass of the intermediate squark/gluino propagator becomes
\bea
 s_{12}&=&2p_{T_1}p_{T_2}(\cosh\eta_1\cosh\eta_2-\sinh\eta_1\sinh\eta_2-
 \cos\phi_2),
\eea
so that
\bea
 \delta(s_{12}-m_{\sq,\gl}^2)&=&{\delta\left[\phi_2-\arccos\left(\cosh\eta_1
 \cosh\eta_2-\sinh\eta_1\sinh\eta_2-{m_{\sq,\gl}^2\over2p_{T_1}p_{T_2}}
 \right)\right] \over p_{T_1}p_{T_2}|\sin\phi_2|}. 
\eea
The three-body cross section
\bea
 \d\sigma& \!=\! &\int
 x_a f_{a/A}(x_a,M_a^2)
 x_b f_{b/B}(x_b,M_b^2)
 {\d\hat{\sigma}\over\d t}
 {\rm BR}(\sq,\gl\to\gr X)
 {2\over\pi|\sin\phi_2|}
 \d p_{T_1} \d\eta_1
 \d p_{T_2} \d\eta_2
 \d\eta_3~~~
 \label{eq:23}
\eea
can then be expressed in terms of the squared and averaged $2\to2$
production cross section of gravitinos and squarks or gluinos,
d$\hat{\sigma}/$d$t$ (see Sec.\ \ref{sec:2a}), and the squark or gluino
branching ratio BR$(\sq,\gl \to\gr X)$ into a gravitino and a jet (see
Sec.\ \ref{sec:2b}).

The transverse-momentum spectrum of the observed jet, which is equivalent
to the missing transverse-momentum spectrum, is shown in Fig.\ \ref{fig:9}
%
\begin{figure}
 \centering
 \includegraphics[width=0.9\columnwidth]{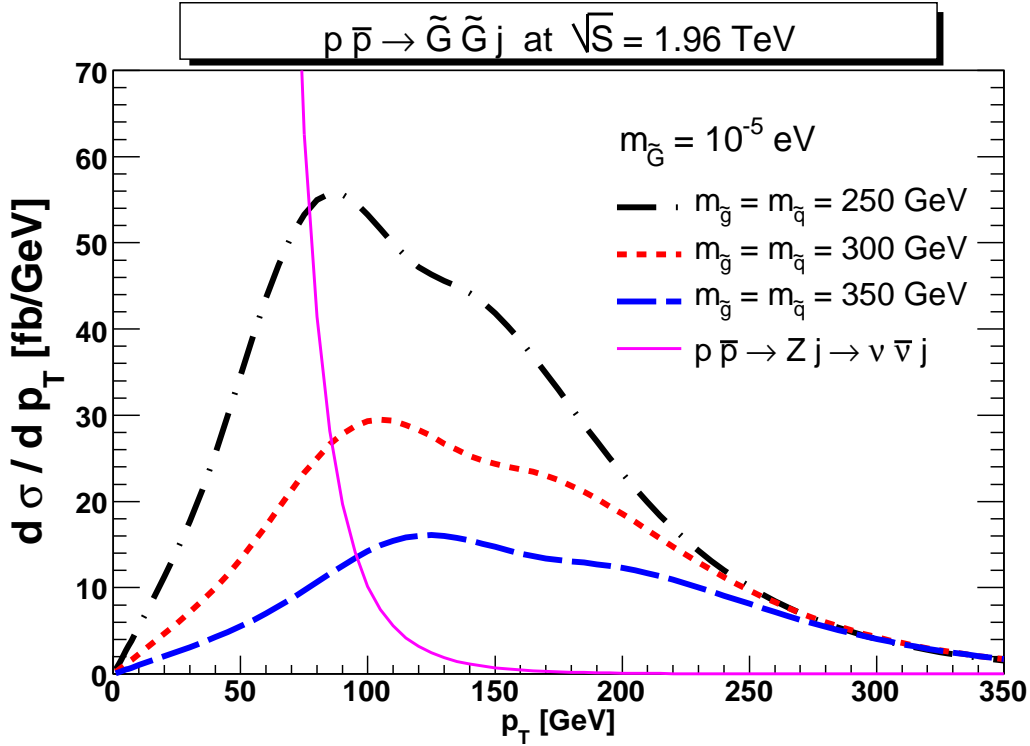}
 \caption{\label{fig:9}Transverse-momentum spectra of the observed jet in
 gravitino production at Run II of the Tevatron for $\mr=10^{-5}$ eV and
 three different squark/gluino masses. Also shown is the main SM background
 from invisible $Z$-boson decays.}
\end{figure}
%
for the same gravitino mass of 10$^{-5}$ eV as in Fig.\ \ref{fig:8}. At this
point, squarks and gluinos always decay into gravitinos and jets (see Fig.\
\ref{fig:6}). Their decay widths vary between 4 and 17\% of their masses, as
these increase from the current exclusion limit of 250 GeV to 350 GeV, so
that the narrow-width approximation is always justified. All three spectra
peak at values slightly below half of the squark/gluino mass, as expected
from kinematic considerations. The main SM background, which comes from the
associated production of a jet and a $Z$-boson, followed by an invisible
$Z$-decay, peaks at roughly half the $Z$-boson mass. It can be eliminated by
cutting on the invisible (or jet) transverse momentum at values around 100
GeV.

In a recent analysis of Run I Tevatron data, the CDF collaboration have
examined events with a single jet and a missing transverse energy of at
least 100 GeV \cite{Affolder:2000ef}. From Fig.\ \ref{fig:9} it is clear
that, while this cut eliminates basically all of the soft-QCD and other
Standard Model backgrounds, only little signal cross section is lost.
For an optimized missing transverse energy cut of 175 GeV, the CDF
collaboration found an upper limit of the gravitino cross section of
3.1 pb, corresponding to a gravitino mass of at most $1.1 \cdot 10^{-5}$ eV.
Note however that this analysis was done under the assumption that all other
supersymmetric particles are heavy \cite{Brignole:1998me}.

We therefore repeat the CDF analysis for general SUSY scenarios, using the
95\% confidence limits on the product of the acceptance ($A$) times the
signal cross section $(\sigma)$ as a function of missing-$E_T$ as published
in Fig.\ 3 of Ref.\ \cite{Affolder:2000ef}. These limits are divided by the
detector acceptance $(A=0.4)$ for the selected data sample, assumed to have
little dependence on the missing-$E_T$ \cite{castro:private}. The dependence
of the number of simulated signal events on the varying missing-$E_T$ cut is
then taken into account explicitly by integrating Eq.\ (\ref{eq:23}) over
$p_{T_1}\geq100~...~300$ GeV. We also implement the experimental requirement
that at least one jet lies in the central region, $|\eta|\leq 0.7$, since
this cut is stricter than the additional CDF-cut on the hardest jet to lie
in $|\eta|\leq2.4$, and our parton-level analysis has only one jet. We have
verified that the number of events with a hard jet in the region $0.7\leq|
\eta|\leq2.4$ is indeed negligible.

From our confirmation of Figs.\ 5 and 6 in \cite{Kim:1997iw} we know that
the two production processes that involve both the squark and gluino mass
($q\bar{q}\to\gr\gl$ and $qg\to\gr\sq$) are bounded from below for $\ms=
\mg$, while the third production process $gg\to\gr\gl$ depends only on the
gluino mass. We sum therefore over all three production subprocesses with
$\ms=\mg$ and take BR $=$ 1 for $\mr\leq10^{-4}$ eV according to Fig.\
\ref{fig:6}.

\begin{figure}
 \centering
 \includegraphics[width=0.9\columnwidth]{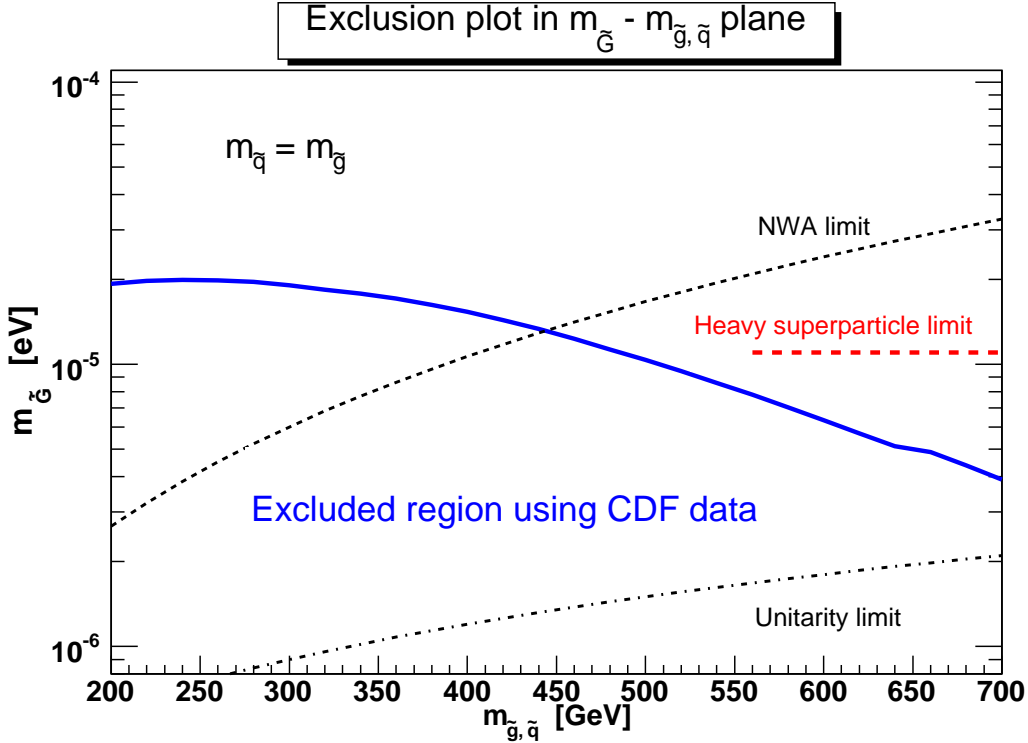}
 \caption{\label{fig:10}Exclusion contour (full curve) in the $\mr-m_{\gl,
 \sq}$ plane derived from the CDF acceptance times cross section limits for
 events with a single jet and varying missing transverse energy
 \cite{Affolder:2000ef,castro:private}. Also shown are the validity region
 of the narrow-width approximation (NWA, above the dotted curve), the CDF
 limit obtained for very heavy squarks and gluinos (dashed line), and the
 unitarity limit (dot-dashed line).}
\end{figure}

By imposing always the strongest experimental experimental limit on $A\,
\sigma\,(\not{\!\!E}_T)$ of Fig.\ 3 in Ref.\ \cite{Affolder:2000ef}, divided
by $A=0.4$ \cite{castro:private}, on the correspondingly integrated cross
section, Eq.\ (\ref{eq:23}), we obtain a contour in the $\mr-m_{\gl,\sq}$
plane, which is shown in Fig.\ \ref{fig:10} (full curve). For light squark
%
%
and gluino masses of 200 GeV, we find a gravitino mass limit of $2\cdot
10^{-5}$ eV that is very similar to that found by CDF for very heavy squarks
and gluinos (dashed line). The limit in Fig.\ \ref{fig:10} degrades slowly
to $4\cdot10^{-6}$ eV as the squark/gluino mass increases to 700 GeV, i.e.\
as it approaches the center-of-mass energy available at the Tevatron and the
theoretical cross section falls. At the same time, the squark/gluino width
increases and eventually passes the value of 1/4 of the squark/gluino mass
(dotted curve in Fig.\ \ref{fig:10}). Our results, obtained in the
narrow-width approximation for single-gravitino production in association
with relatively light squarks and gluinos, are thus complementary to those
obtained by CDF for very heavy squarks and gluinos and double-gravitino
production \cite{Affolder:2000ef}. We have also checked that for our
analysis, which is based on a single-gravitino effective Lagrangian,
tree-level unitarity is always satisfied \cite{Bhattacharya:1988nk}, since 
\bea
 {\mr\over10^{-6}~{\rm eV}}&\geq&0.3\,{\mg\over100~{\rm GeV}}
\eea
for a critical energy corresponding to the Tevatron center-of-mass energy
(dot-dashed curve in Fig.\ \ref{fig:10}).

Two other experimental analyses of monojet signals at the Tevatron have been
published, one based on 78.8 pb$^{-1}$ of Run-I data by the D0 collaboration
\cite{Abazov:2003gp} and one based on 368 pb$^{-1}$ of Run-II data by the
CDF-collaboration \cite{Abulencia:2006kk}. However, both analyses are
interpreted with extra-dimensional models and directly present limits on the
number of these extra dimensions and on the corresponding fundamental Planck
scale. It would be interesting to re-interpret these analyses in the context
of gravitino production. The CDF analysis quotes indeed a model-independent
limit on signal events (or signal cross section times acceptance), but
neither publication gives numerical values for detector acceptances. These
were also not available from the collaborations upon request, so that we
can at this point not deduce independent gravitino mass limits from these
data.

\subsection{LHC}

\begin{figure}
 \centering
 \includegraphics[width=0.9\columnwidth]{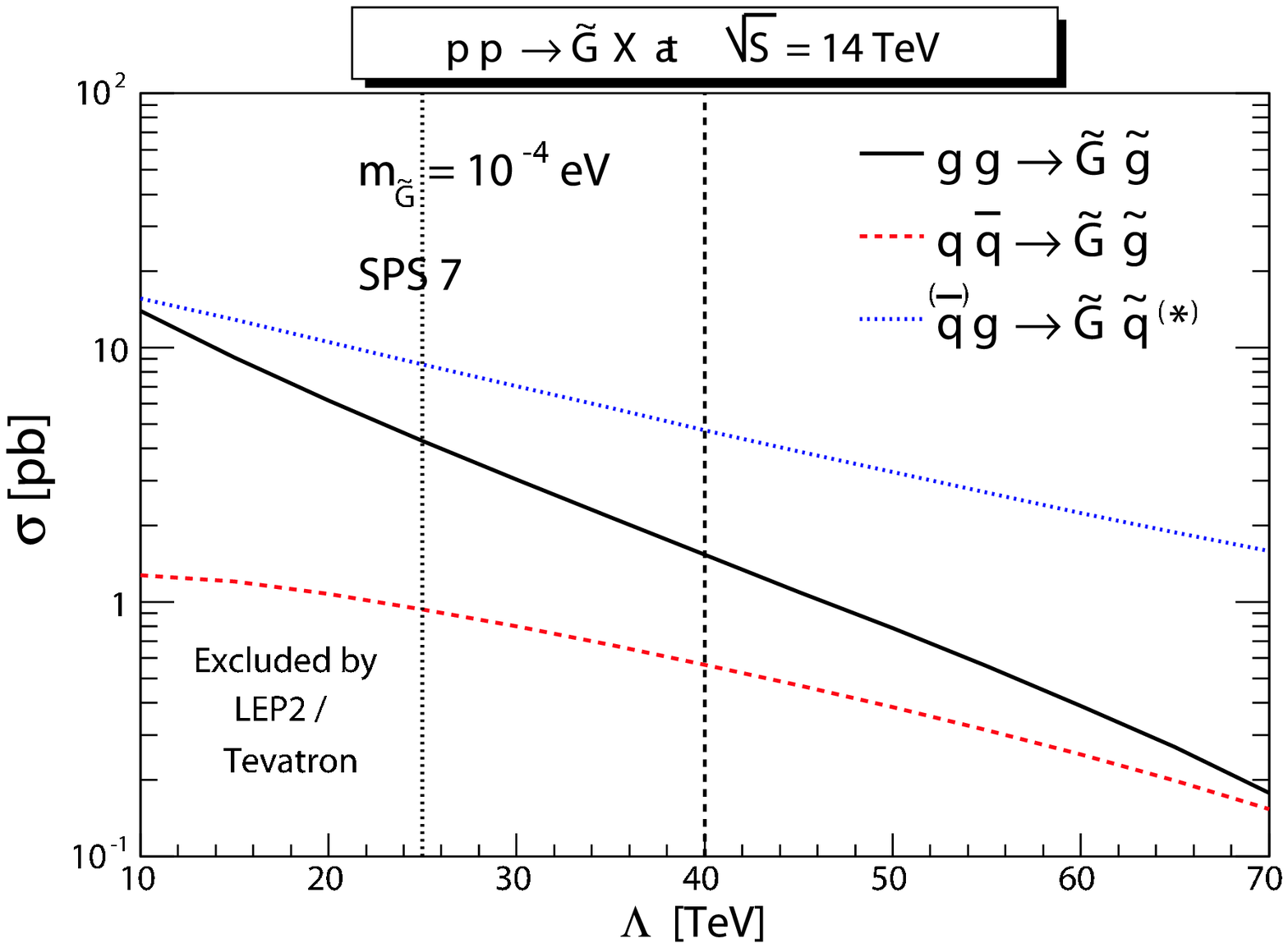}
 \includegraphics[width=0.9\columnwidth]{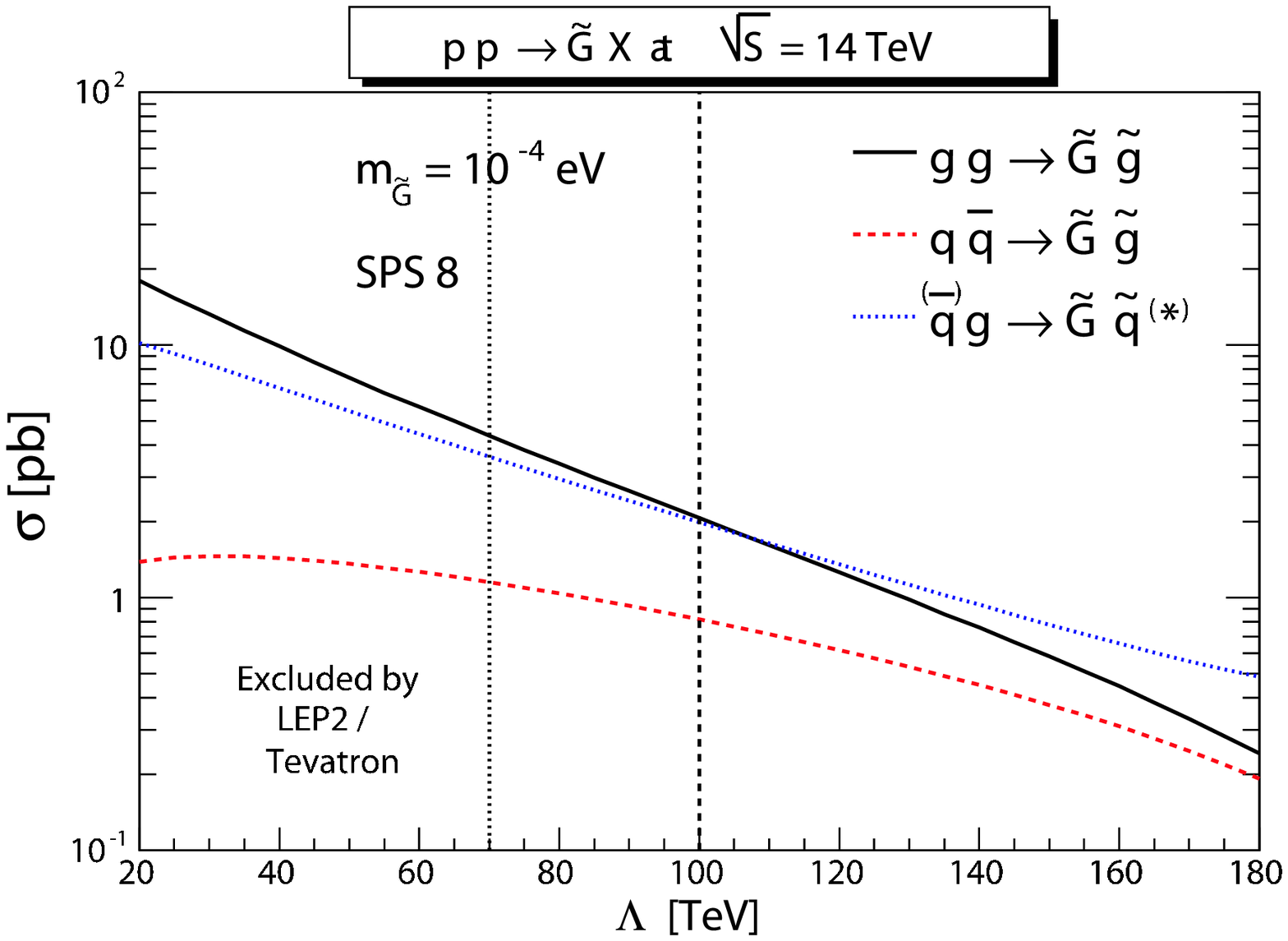}
 \caption{\label{fig:11}Total cross sections of gravitino and gluino/squark
 associated production at the LHC for $\mr=10^{-4}$ eV and the GMSB
 benchmark slopes SPS 7 (top) and SPS 8 (bottom) as a function of the
 effective SUSY-breaking scale $\Lambda$.}
\end{figure}

The high center-of-mass energy of $\sqrt{S}=14$ TeV and the large luminosity
of initially 10 fb$^{-1}$ and finally 300 fb$^{-1}$ available at the LHC
will provide the opportunity to test the soft SUSY-breaking hypothesis up
to the multi-TeV range. This general remark remains true for the
gauge-mediated SUSY-breaking scenarios SPS 7 (top) and 8 (bottom) with
a gravitino LSP that we consider in Fig.\ \ref{fig:11}. In this figure, we
%
%
show the total cross sections of gravitino and gluino/squark associated
production at the LHC for $\mr=10^{-4}$ eV and the three different partonic
gravitino production processes discussed above. Their hierarchy is now
opposite to the one at the Tevatron, i.e.\ it is the gluon luminosity
that dominates and no longer the quark-antiquark luminosity. At SPS 7,
where $\overline{m}_{\sq}\leq\mg$, squarks are produced more copiously than
gluinos, wheres the inverse is true at SPS 8. In both cases, the discovery
reach extends to values of the effective SUSY-breaking scale $\Lambda$ far
above the actual benchmark points. It is clear that a striking monojet
signal with large missing transverse energy could be discovered rapidly
after the start-up of the LHC and with rather low luminosity.
The only existing previous analysis for gravitino production at the LHC,
based on the dominating subprocess $gg\to\gr\gl$ only, assumed a slightly
higher LHC center-of-mass energy of $\sqrt{S}=16$ TeV and fixed squark and
gluino masses of 2 TeV and 750 GeV, respectively \cite{Drees:1990vj}. The
authors computed a monojet cross section of similar size as the one shown
in the lower part of Fig.\ \ref{fig:11} (8 pb for $\mr=10^{-4}$ eV) for
comparable squark and gluino masses.

\begin{figure}
 \centering
 \includegraphics[width=0.9\columnwidth]{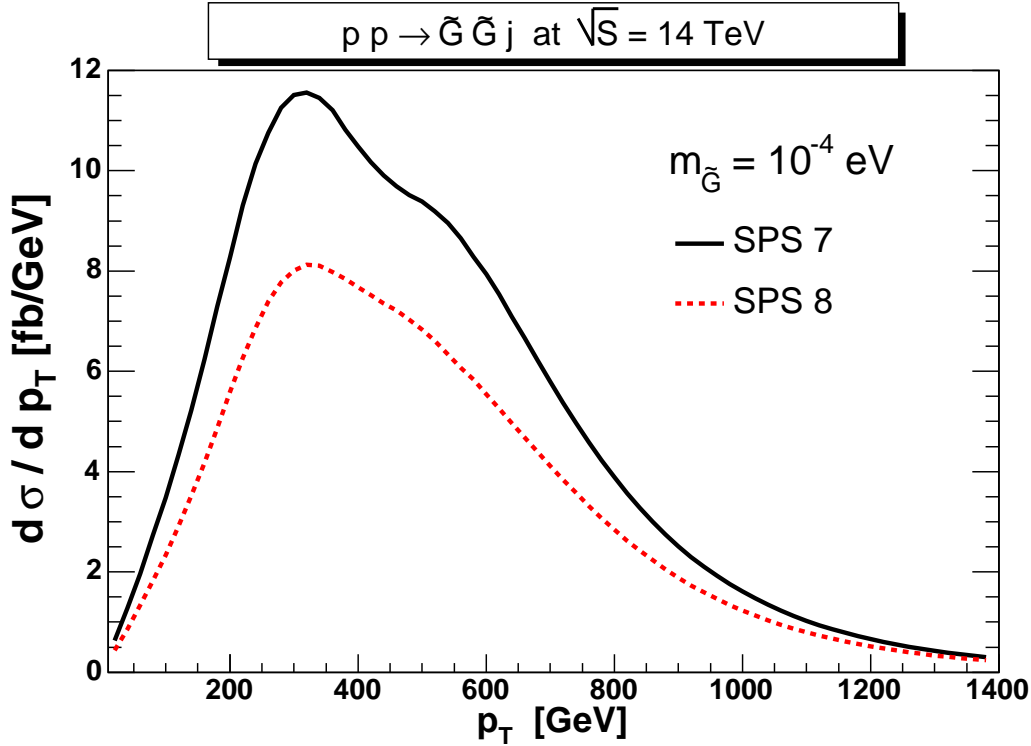}
 \caption{\label{fig:12}Transverse-momentum spectra of the observed jet in
 gravitino production at the LHC for SPS 7 (full line) and SPS 8 (dashed
 line).}
\end{figure}

The transverse-momentum spectra at the LHC are shown in Fig.\ \ref{fig:12}
for the two GMSB benchmark points and a gravitino mass of $\mr=10^{-4}$ eV.
%
%
While the spectra peak again at values slightly below half of the
squark/gluino masses, as was already the case at the Tevatron (see Fig.\
\ref{fig:9}), they extend now to much larger values of $p_T\simeq 1200$ GeV.
Note also that the absolute Tevatron cross section was only of similar size
since it had been calculated with a smaller gravitino mass of
$\mr=10^{-6}$ eV.
In the high-luminosity phase at the LHC, the missing transverse-energy
signal will be degraded by pile-up events. In the ATLAS experiment, pile-up
can be eliminated and acceptable trigger rates in the kHz-range can be
obtained for missing-$E_T$ thresholds of 60/120 GeV at low/high luminosity.
These values can be two times lower if an addition hard jet of $E_T>100$ GeV
is required \cite{Wunsch:1998}. As can be seen from Fig.\ \ref{fig:12}, the
signal cross section will be affected very little by these cuts.

\section{Conclusion}
\label{sec:4}

In this paper, we have derived the Feynman rules for light gravitino
production and decay from an effective supergravity Lagrangian in
four-component, non-derivative form and computed analytical partonic cross
sections and branching ratios involving interactions of gravitinos, gluinos,
and squarks. Special emphasis has been put on the gauge-independence of the
results, the contributions of quartic vertices, and a comparison with
results obtained previously with a derivative Lagrangian and those obtained
in the high-energy limit.

Using the narrow-width approximation, we combined the associated
gravitino-squark/gluino production cross sections with the subsequent decay
of the squarks/gluinos into quarks/gluons and a second gravitino. This
enabled us to perform extensive numerical studies of branching ratios, of
the total hadronic cross sections at the Tevatron and the LHC, and of the
corresponding transverse-energy spectra of the single observed jet, then to
impose experimental cuts on the latter, and finally to derive a new and
robust exclusion contour in the $\mr/m_{\sq,\gl}$-plane from the latest CDF
monojet cross section limit.

Our Tevatron exclusion contour implies that gravitinos with masses below
$2\cdot10^{-5}$ to $1\cdot10^{-5}$ eV are excluded for squark/gluino-masses
below 200 and 500 GeV, respectively. These limits are complementary to the
one obtained by the CDF collaboration, $1.1\cdot10^{-5}$ eV, obtained under
the assumption of very heavy squarks and gluinos.

For the LHC, we conclude that SUSY scenarios with light gravitinos, such as
the GMSB benchmark slopes SPS 7 and 8, will lead to a striking monojet
signal very quickly after its startup in 2008 and already with low\
luminosity. The missing-$E_T$ and jet trigger thresholds foreseen by the
ATLAS collaboration are perfectly suitable also in these scenarios for an
efficient background reduction without affecting the signal in a significant
way.

\section*{Acknowledgments}
M.K.\ gratefully acknowledges the hospitality of the Institute of
Theoretical Physics at the University of G\"ottingen, where part of
this work has been completed.
%

\appendix

\section{Effective Lagrangian for Single Goldstinos}
\label{sec:a}

We start from the effective Lagrangian in four-component notation for the
single interaction of a light gravitino \cite{Fayet:1986zc,Moroi:1993mb,%
Gherghetta:1996fm,Giudice:1998bp}, whose longitudinal (spin-1/2) goldstino
components \cite{Casalbuoni:1988kv} can couple to the matter and gauge
supermultiplets with enhanced (electroweak) strength \cite{Fayet:1977vd}.
The corresponding effective Lagrangian in two-component notation can be
found in \cite{Clark:1996aw,Brignole:1996fn,Brignole:1997pe,Luty:1998np,%
Clark:1997aa,Lee:1998xh,Lee:1998aw,Brignole:1999gf}, while interactions
involving several external goldstinos or goldstino propagators and the
scalar/pseudo-scalar superpartners of the goldstinos, the so-called
sgoldstinos, have been derived in four-component notation, e.g., in
\cite{Gherghetta:1996fm,Brignole:1997sk}.

By correctly translating the effective Lagrangian for light gravitinos in
two-component notation \cite{Lee:1998xh,Lee:1998aw} into the four-component
notation of the traditional SUSY-QCD Lagrangian \cite{Haber:1984rc}, we
obtain the following non-derivative couplings for the interactions of
Majorana-fermionic goldstinos $\psi$ with Dirac-fermionic quarks $\chi$,
complex-scalar squarks $\phi$, vector-bosonic gluons $A_\mu^a$, and
Majorana-fermionic gluinos $\lambda^a$:
\bea
 {\cal L}^{\rm eff}&=&{\ms^2-m_q^2\over\sqrt{3}M\mr}
 \lr\overline{\chi}P_L\psi\,\phi_R
   -\overline{\chi}P_R\psi\,\phi_L
   +\overline{\psi}P_R\chi\,\phi_R^*
   -\overline{\psi}P_L\chi\,\phi_L^*\rr
 \nonumber\\
 &+&{i\mg\over4\sqrt{6}M\mr}\overline{\psi}\,[\gamma^\mu,\gamma^\nu]\,
 \lambda^a F^a_{\mu\nu}
 -{g_s\mg\over\sqrt{6}M\mr}\,\overline{\psi}\lambda^a\phi_i^*T^a_{ij}
 \phi_j.\label{eq:1}
\eea
Here, $M=(8\pi G_N)^{-1/2}=2.435 \cdot 10^{18}$ GeV is the reduced
Planck mass, $\mr$ is the gravitino mass, which is related to the
supersymmetry breaking vacuum expectation value $\langle F \rangle$ in
canonical normalization by $\mr=\langle F \rangle/(\sqrt{3}M)$, and $g_s$ is
the strong gauge coupling. $T^a_{ij}$ are the generators of the $SU(N_C=3)$
color symmetry group with antisymmetric structure constants $f^{abc}$ and
Casimir operator $C_F=4/3$, and $P_{L,R}=(1\mp\gamma_5)/2$ are the
chirality projection operators. The squark and gluino masses will be denoted
$\ms$ and $\mg$. Since the top quark density in hadrons is small, we can
neglect the masses $m_q$ of the five light quarks at high collision energies
and consider the corresponding left- and right-handed squarks to be
mass-degenerate.

The effective theory contains the same couplings as the full theory
\cite{Cremmer:1982en,Wess:1992cp}, except for the
quark-gluon-gravitino-squark vertex, which would violate the gauge symmetry
in the effective theory \cite{Lee:1998aw}, but was erroneously kept in the
alternate Feynman rules of \cite{Kim:1997iw}. These were, however, not used
there to calculate cross sections. In contrast, there is a new four-particle
vertex \cite{Lee:1998xh,Lee:1998aw}, the gravitino-gluino-squark-squark
vertex, which has been overlooked in all other cited references, but is
neither relevant for our analysis nor for the one in \cite{Kim:1997iw}.
Attention must also be paid to the assignment of factors of $i$ in Eq.\
(\ref{eq:1}), if the interference terms in \cite{Kim:1997iw,Lopez:1996ey}
between scalar and gauge boson exchanges are to be correctly reproduced.

In the effective Lagrangian, all vertices are proportional to
SUSY-breaking mass terms, i.e.\ $\ms^2-m_q^2$ and $\mg$. In particular,
the Yukawa coupling of the goldstino can be obtained from that of the
gluino by the replacement \cite{Fayet:1986zc}
\bea
 g_s T^a_{ij}&\to&{\ms^2-m_q^2\over\sqrt{6}M\mr},
\eea
while the goldstino-gluon-gluino coupling can be obtained by the
replacement
\bea
 -g_sf^{abc}\gamma^\mu&\to&i\,{\mg\over2\sqrt{6}M\mr}\,\delta^{ab}\,
 [\not{\!\!P},\gamma^\mu]
\eea
with $P$ representing the incoming gluon-momentum. At high energies,
contributions involving the cubic goldstino-quark-squark coupling are
suppressed relative to the gluino contribution due to the higher
mass-dimension of the coupling.

\section{Feynman Rules for Single Goldstinos}
\label{sec:b}

In order to derive the Feynman rules needed for the hadronic production and
decay of single goldstinos, we multiply the effective Lagrangian in Eq.\
(\ref{eq:1}) with a factor of $i$, perform a Fourier transformation, and
take the functional derivative with respect to the external fields.
Denoting the incoming four-momentum by $P$, the chirality projection
operators by $P_{L,R}=(1\mp\gamma_5)/2$, Lorentz indices by $\mu,\,\nu,\,
...$, and color indices of the fundamental (adjoint) representation of the
color symmetry group SU(3) by $i,\,j,\,...$ $(a,\,b,\,...)$, we obtain the
following interaction vertices:
%
\bea
 \epsfig{file=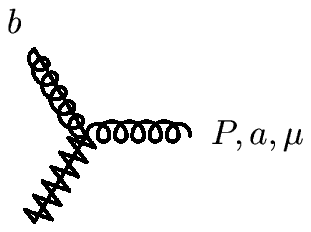,width=4cm} &\hspace*{15mm}&
 +\,i\,{\mg\over2\sqrt{6}M\mr}\,\delta^{ab}\,[\not{\!\!P},\gamma^\mu] \\
 \epsfig{file=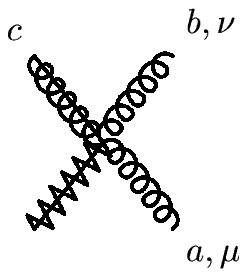,width=4cm} &&
 +\,{\mg\over2\sqrt{6}M\mr}\,g_s\,f^{abc}\,[\gamma^\mu,\gamma^\nu] \\
 \epsfig{file=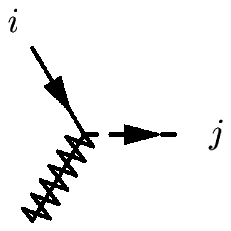,width=4cm} &&
 \mp\,i\,{\ms^2-m_q^2\over\sqrt{3}M\mr}\,\delta_{ij}\,P_{L,R} \\
 \epsfig{file=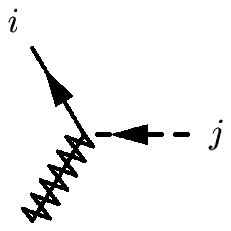,width=4cm} &&
 \mp\,i\,{\ms^2-m_q^2\over\sqrt{3}M\mr}\,\delta_{ij}\,P_{R,L} \\
 \epsfig{file=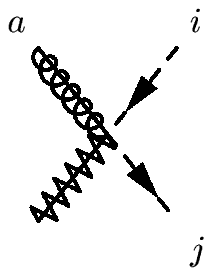,width=4cm} &&
 -\,i\,{\mg\over\sqrt{6}M\mr}\,g_s\,T^a_{ij}\\ \nonumber
\eea
%

Here, the arrows on (s)quark lines indicate flavor flow, while the Majorana
nature of gravitinos and gluinos requires the fermion flow to be fixed
arbitrarily \cite{Denner:1992vz}. These Feynman rules have been implemented
into the computer algebra program {\tt FeynArts} \cite{Kublbeck:1990xc,%
Hahn:2001rv}, and the corresponding model file is available from the authors
upon request. The derivative forms of the (s)goldstino interaction vertices
in two-component form have already been implemented some time ago into the
program {\tt CompHEP} \cite{Gorbunov:2001pd}.

Our Feynman rules differ from those presented in appendix A.3.3 of
\cite{Bolz:2000xi}, which have also been derived from \cite{Lee:1998aw}, by
a factor of $i$ in the first two vertices, apparently due to a
misinterpretation of the two-component tensor $\sigma^{\mu\nu}$
\cite{Lee:Private}. The Feynman rules in \cite{Moroi:1995fs} differ from
ours in addition by an irrelevant global factor of $i$.
The usual SUSY-QCD vertices and propagators can be found, e.g., in
\cite{Hopker:1996sx}. 



\end{document}